\documentclass[aps,prl,twocolumn,groupedaddress,amsmath,amssymb,noeprint]{revtex4-2}
\usepackage{graphicx}
\usepackage{bm}
\usepackage{xr}
\makeatletter
\usepackage[colorlinks=True]{hyperref}
\usepackage{cleveref}
\begin{document}
\title{
Exploiting memory effects to detect the boundaries of biochemical subnetworks }%

\author{Moshir Harsh}
\email{moshir.harsh@uni-goettingen.de}

\author{Leonhard G{\"o}tz Vulpius}

\author{Peter Sollich}
\email{peter.sollich@uni-goettingen.de}

\altaffiliation[also at\ ]{Department of Mathematics, King’s College London, London WC2R 2LS, UK}
 
\affiliation{Institut f{\"u}r Theoretische Physik\\ Georg-August-Universit{\"a}t G{\"o}ttingen, Germany}

\date{\today}

\begin{abstract}
Partial measurements of biochemical reaction networks are ubiquitous and limit our ability to reconstruct the topology of the reaction network and the strength of the interactions amongst both the observed and the unobserved molecular species. Here, we show how we can utilize noisy time series of such partially observed networks to determine which species of the observed part form its boundary, i.e.\ have significant interactions with the unobserved part. This opens a route to reliable network reconstruction. The method exploits the memory terms arising from projecting the dynamics of the entire network onto the observed subnetwork. We apply it to the dynamics of the Epidermal Growth Factor Receptor (EGFR) network and show that it works even for substantial noise levels.
\end{abstract}

\maketitle

Inference of the topology and interactions of biochemical reaction networks from noisy dynamical measurements is a problem of prime importance for the data-driven elucidation of their structure and biological function. Methods tackling this apply more  broadly to complex networks~\cite{barabasi_taming_2005,strogatz_exploring_2001,boccaletti_complex_2006} in e.g.\ climate~\cite{donges_complex_2009} or power systems~\cite{pagani_power_2013}.

Amongst various proposed methods~\cite{timme_revealing_2014} a common approach~\cite{bonvin_target_1990,burnham_inference_2008, willis_inference_2016, han_robust_2015,pan_identifying_2015,casadiego_model-free_2017,mangan_inferring_2016} models the dynamics using {\em a priori} kinetic knowledge to represent inter-species interactions. Another class uses dynamical correlations~\cite{barzel_network_2013,ching_reconstructing_2015,ching_reconstructing_2017} or information theoretical methods~\cite{braunstein_inference_2008,quinn_estimating_2011} to reconstruct network connectivity. Yet other methods~\cite{nitzan_revealing_2017,timme_revealing_2014} rely on the system's response to external driving. These methods are, however, not generally designed to handle the very common scenario where only parts of a much larger network can be observed, and when applied in such scenarios they perform significantly worse~\cite{swain_intrinsic_2002,kaern_stochasticity_2005}. A noteworthy proposal~\cite{langary_inference_2019} reconstructs the reaction stoichiometry without rate constants and can, in special cases, tolerate unobserved species by effectively ignoring them. Other approaches that exploit first-passage time distributions~\cite{li_mechanisms_2013,valleriani_unveiling_2014} to count intermediate states require high temporal resolution data and are unsuitable for reaction networks with their exponentially large state spaces.

In this letter, we propose a simple yet efficient data-driven approach that, in a partially observed network -- the {\em subnetwork} -- identifies those molecular species that lie on its {\em boundary} in the sense that they have significant interactions with the unknown part of the network (the {\em bulk}). Such an identification can then form the basis for systematic expansion of the observed network by exploring the dynamics and potential interaction partners of the boundary species. Our work thus poses the question of how the topology of a biochemical subnetwork relative to an embedding bulk network can be elucidated, and provides a method that begins to address this challenge.

Our approach uses Bayesian regression and exploits recent insights~\cite{rubin_memory_2014,herrera-delgado_memory_2018,bravi_systematic_2020} into the boundary structure of memory in the projected dynamical equations that describe the non-equilibrium dynamics of a subnetwork. (In contrast, techniques~\cite{tang_dynamical_2020,haehne_detecting_2019,borner_revealing_2023} that estimate total network {\em size} have limited utility for our problem because they provide no information about the interactions of the subnetwork with the bulk.) The inputs to our method are noisy, non-steady-state relaxation time traces of the observed subnetwork, without requiring any external driving. Below we explain the approach in general terms and then characterise it on a simple example network and a real biological example.

\textbf{Reaction network and dynamics}. A chemical reaction network consists of $P$ different chemical species $X_i$, $i \in \{1,\dots,P\}$ that react in $R$ reactions. It is specified by the entries $S_{ir}$ of the stoichiometric matrix $\bm{S}$, which give the net change in copy number of species $i$ in reaction $r$, and the associated rate constants $k_r$. In the limit of large reaction volume $V$ we work with continuous concentration variables $\bm{x}^*=\bm{n}/V=(x_1,\dots,x_P)^{\rm T}$ where the vector $\bm{n}$ contains the number of molecules of each species.
We assume mass action reaction propensities~\cite{schnoerr_approximation_2017,waage_studier_1864,harsh_see_nodate}; to account for fluctuations around the average concentrations to leading order in inverse reaction volume $\epsilon = V^{-1}$, we consider reaction kinetics described by the Chemical Langevin Equation~\cite{schnoerr_approximation_2017,harsh_see_nodate}.
\nocite{tipping_sparse_2001,mackay_bayesian_1992,pedregosa_scikit-learn_2011,harsh_accurate_2023,fromberg_generalized_2021}
For $\epsilon = 0$ it reduces to the noise-free case, i.e.\ deterministic mass action kinetics. 

\begin{figure}[hb]
	\includegraphics[width=0.95\linewidth]{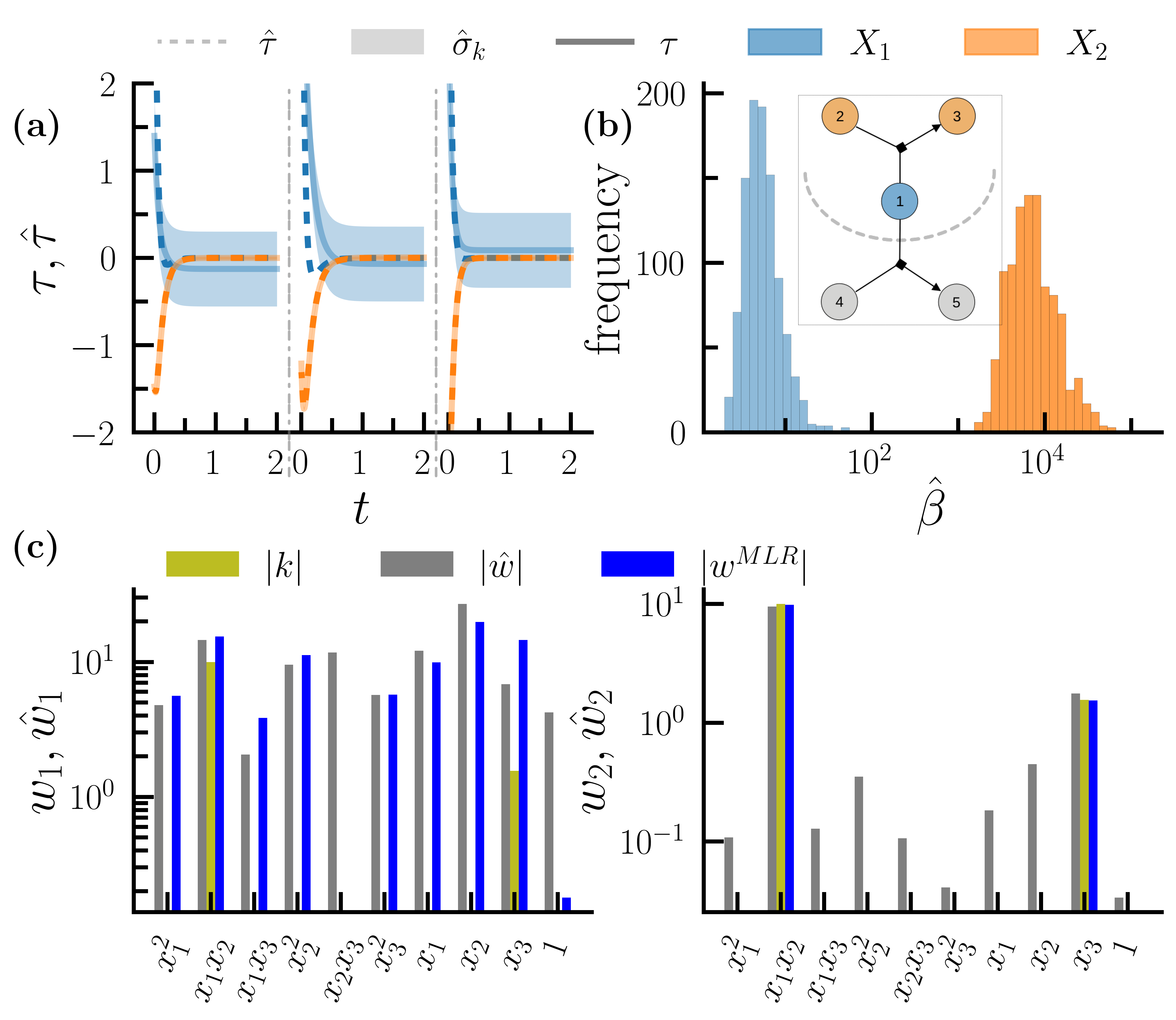}%
	\caption{\label{fig1} (a) Illustration of regression targets, %
		i.e.~concentration time derivatives $\tau$ (solid lines), corresponding estimates $\hat \tau$ (dashed lines) and predictive standard deviations $\hat \sigma$ (shaded areas), for boundary species $X_1$ (blue) and interior species $X_2$ (orange). Three different (numerical) experiments are shown on concatenated time axes. Note that $\hat \sigma_1 \gg \hat \sigma_2$: the boundary species has much higher effective noise.
		(b) Histograms of the inverse effective noise variance $\hat \beta$ from 1000 independent training sets, showing clear separation between 
		$X_1$ and $X_2$; (inset) schematic of the simple five species network. (c) For $X_1$ (left) and $X_2$ (right), absolute values of true rate constants (weights) $k$ for each basis function, estimated weights $\hat w$ from our Bayesian regression and $w^{\rm MLR}$ from a popular %
		multiple linear regression method~\cite{burnham_inference_2008}. }
\end{figure}

\textbf{Subnetwork dynamics}.
Suppose that on account of measurement limitations or because the rest of the network is simply unknown, we can only observe $d$ molecular species, say the $x_i$ with $i =  1,\dots,d$. The dynamics of the concentration vector $\bm{x}$ of the subnetwork species can then be cast in the general form
\begin{equation}   
  \label{eq:theory:zwanzig_mori_general} 
   \partial_t \bm x (t) = \bm W \bm\phi(\bm x(t)) + \int_0^t dt'\,\bm \Gamma(t-t') \bm\phi(\bm{x}(t')) + \bm r(t)
\end{equation} 
using the Zwanzig-Mori projection formalism~\cite{mori_transport_1965,zwanzig_memory_1961}; we focus here on the noise-free case for simplicity. The matrix $\bm{W}$ contains the coefficients of the local-in-time terms; but the projection also leads to memory functions $\bm{\Gamma}$. The final random force term $\bm r(t)$ captures the effects of the unknown initial bulk concentrations~\cite{rubin_memory_2014,bravi_systematic_2020}.

The details of~\cref{eq:theory:zwanzig_mori_general} are governed by the vector of basis functions $\bm{\phi}(\bm{x})$ of the subnetwork concentrations that we project on. We choose $\bm\phi(\bm x)$ as consisting of all monomials in $\bm x$ up to second order, allowing us to represent all reactions up to binary ones via the local-in-time term $\bm{W}\bm{\phi}(\bm{x}(t))$ in \cref{eq:theory:zwanzig_mori_general}. This is in line with previous considerations~\cite{burnham_inference_2008,willis_inference_2016, conradi_subnetwork_2007, herrera-delgado_memory_2018, rubin_memory_2014} and suffices in particular to capture the dynamics of protein interactions~\cite{rubin_memory_2014} and of non-linear enzymatic and gene regulation networks~\cite{herrera-delgado_memory_2018,rubin_michaelis-menten_2016}.

The central observation that we exploit is the {\em boundary structure} of the memory~\cite{rubin_memory_2014,herrera-delgado_memory_2018}. For the {\em interior} subnetwork species, i.e.\ those that have no interactions with the bulk, the memory term and random force both disappear and \cref{eq:theory:zwanzig_mori_general} reduces to the first term on the r.h.s., with $\bm{W}$ containing the rate constants of the relevant subnetwork reactions. For the {\em boundary} species, on the other hand, nonzero memory and random force terms are present. We rephrase this in terms of a prediction or {\em regression} problem: the concentration changes $\tau_i \equiv \partial_t x_i$ for interior species should be predictable with high accuracy from the subnetwork concentrations $\bm{x}$ at that time, while this will not be possible for the boundary species where memory and random force terms will generate effective noise. Our strategy will therefore be to estimate the effective noise variance in the prediction of each $\partial_t x_i$ and use this to discriminate boundary and interior species. As a by-product we will obtain estimates for the subnetwork reaction rate constants and thence the subnetwork topology.

\textbf{Bayesian Regression}.
Assume that we have a time series, possibly irregularly spaced, of subnetwork concentrations $\bm{x}$. From each time interval we extract an estimate of the concentration time derivative $\tau_i = \partial_t x_i$. This is our target for prediction on the basis of the subnetwork concentrations $\bm{x}$ at the beginning of the time interval. Data from $N$ time intervals, possibly concatenated from multiple time series, gives us $N$ inputs $\bm \Xi = (\bm x^{(1)}, \bm x^{(2)},\dots, \bm x^{(N)})$ and $N$ outputs $\bm \tau_i = (\tau_i^{(1)}, \dots,\tau_i^{(N)})^{\rm T}$ whose relation we model as in \cref{eq:theory:zwanzig_mori_general} by
\begin{equation}  
  \tau_i^{(n)} = \bm{w}_i^{\rm T}\bm\phi(\bm{x}^{(n)}) + \zeta_i^{(n)}
 \label{eq:regression}
\end{equation} 
with $n=1,\dots,N$ and Gaussian noise $\zeta_i^{(n)}$ of variance $1/\beta_i$. As explained we choose as basis functions the concentration monomials up to second order, $\bm \phi(\bm x) = (1, x_1,\dots,x_d,x_1 x_1,x_1 x_2,\dots,x_d x_d)^{\rm T}$.

To estimate $\beta_i$ we use a Bayesian model selection approach with a zero mean Gaussian prior $p(\bm{w}_i|\alpha_i)$ on the weights $\bm{w}_i$ with covariance matrix $\alpha_i \mathbb{I}$. Combining with the Gaussian likelihood implied by \cref{eq:regression} and integrating over $\bm{w}_i$ gives the so-called marginal likelihood $p(\bm{\tau}_i| \bm{\Xi},\alpha_i,\beta_i)$, which can be expressed in terms of the design matrix $\bm\Phi$ with elements $\Phi_{nm} = \phi_m(\bm x^{(n)})$~\cite{harsh_see_nodate}. The posterior distribution of the model parameters is then, from Bayes' theorem,
\begin{equation}{\label{eq:posterior}}
p(\alpha_i,\beta_i | \bm\tau_i, \bm\Xi) \propto p(\bm{\tau}_i| \bm{\Xi},\alpha_i,\beta_i)p(\alpha_i,\beta_i)
\end{equation}
where the last factor is the (hyper-)prior on $(\alpha_i,\beta_i)$. Taking this to be flat, the model with the highest posterior probability is obtained by maximizing~\cite{harsh_see_nodate} the marginal likelihood w.r.t.\ $\alpha_i$ and $\beta_i$, a procedure called the evidence approximation~\cite{bishop_pattern_2006}, resulting in estimates $\hat\alpha_i$ and $\hat\beta_i$. (Their uncertainties can be obtained from a Gaussian approximation around the peak of the posterior~\cref{eq:posterior}, if desired~\cite{harsh_see_nodate}.) Given $\hat\alpha_i$ and $\hat\beta_i$, the predictive distribution of the weights $\bm{w}_i$ can be calculated in closed form~\cite{harsh_see_nodate}.

On the basis of the $\hat\beta_i$ we then classify the different subnetwork species $X_i$ according to the criterion
\begin{equation}
\label{eq:beta_separation_relation}
    \hat \beta_\text{boundary} \ll \hat \beta_\text{interior}
\end{equation}
based on the expectation that boundary species have effective noise of significantly higher variance $1/\hat\beta_i$.

\begin{figure}
\includegraphics[width=0.95\linewidth]{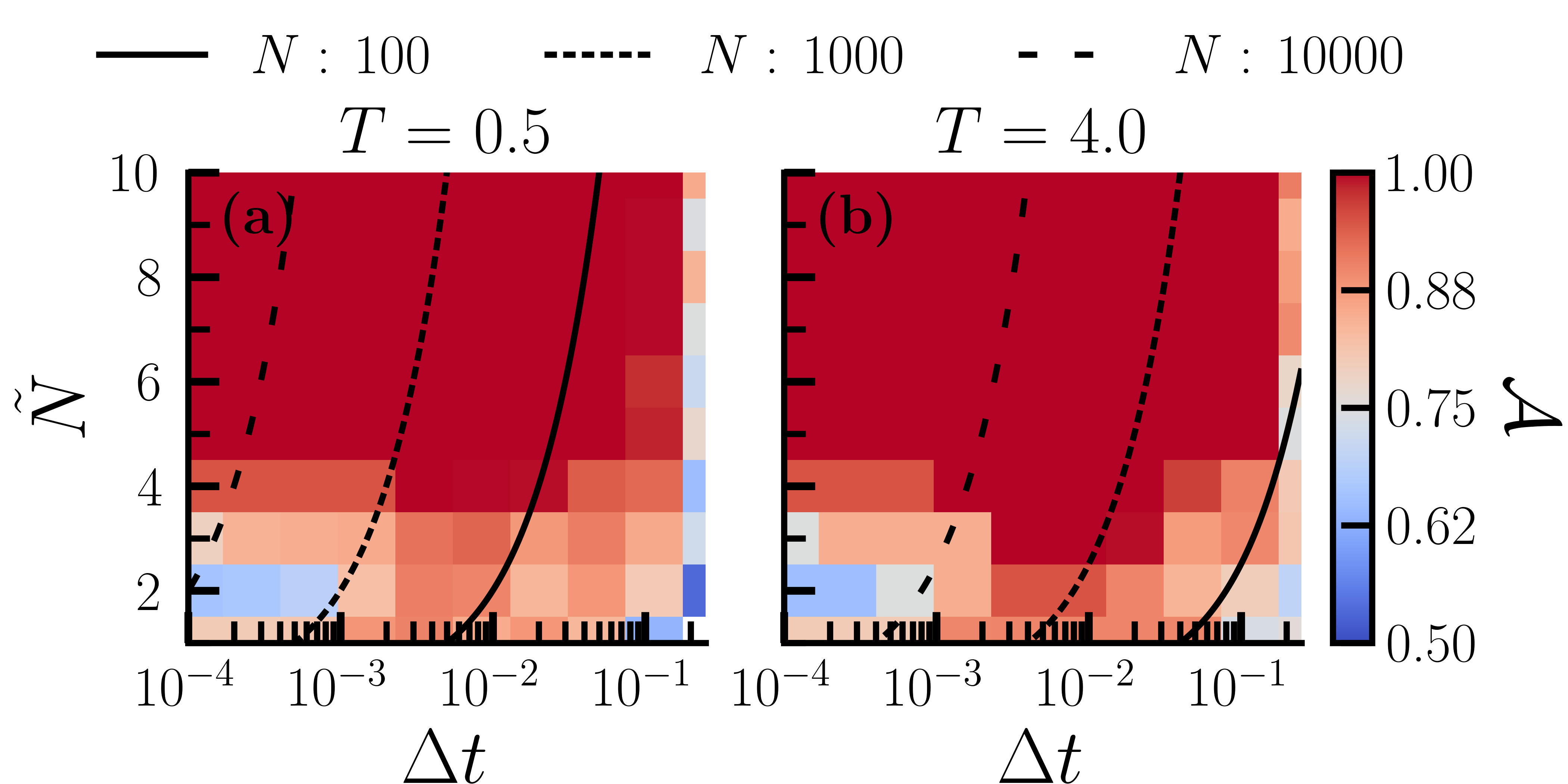}%
\caption{\label{fig3}  Discrimination performance as measured by AUC %
$\mathcal{A}$, as a function of number of experiments $\tilde N$ and time step $\Delta t$, for trajectories of length (a) $T = 0.5 $ and (b) $T=4$, averaged over ten independent data sets for each $(\tilde N,\ \Delta t)$ pair. The number $N$ of training data points is constant along the three lines. Our method works reliably ($\mathcal{A}\approx 1$) for $\tilde N > 4$ and $\Delta t \leq 10^{-2}$. }
\end{figure}

\textbf{Simple Network}. We illustrate the method first for an example network,~\cref{fig1}(b, inset), with five species of which $X_1,X_2,X_3$ form the observed subnetwork. $X_1$ is the boundary species that reacts with the unobserved bulk species $X_4$ and $X_5$. We want to distinguish this from the interior species $X_2$ and $X_3$, given only subnetwork time traces. The reaction dynamics is simulated using the Euler-Maruyama algorithm with boundary reflection~\cite{schnoerr_approximation_2017}, and time series are extracted with a specified time step $\Delta t$ (for parameter values see~\cite{harsh_see_nodate}).

Example results of the regression as displayed in~\cref{fig1}(a) show that the predictive standard deviation of $X_1$ is much larger than for $X_2$ ($X_3$ behaves similarly); correspondingly also the optimized $\hat\beta$ obey $1/\hat\beta_1\gg 1/\hat\beta_2$.  Repeating this for 1000 independent data sets yields a histogram of the $\hat \beta$ (\cref{fig1}(b)) that clearly demonstrates~\cref{eq:beta_separation_relation}: our Bayesian regression can distinguish boundary from interior species. 

The estimated weights for boundary and interior species are plotted in~\cref{fig1}(c) for one of the data sets. The weights corresponding to the true reaction rate constants $k$ are predicted accurately for the interior species while those for non-existent reactions are small. For the boundary species, in contrast, the regression predicts large spurious weights for all basis functions as it tries to represent memory and random force effects in terms of subnetwork reactions. Also shown are results from a popular multiple linear regression (MLR) technique~\cite{burnham_inference_2008} that adds a step of pruning insignificant weight vectors using significance tests while ensuring model consistency.  This, however, fails for the boundary species and does not offer any diagnostic to identify species for which its results are unreliable.

\textbf{Parameters affecting performance}. Important parameters affecting our method's performance are the total length $T$ of each time series, the time step $\Delta t$, and the number $\tilde N$ of time series from different  experiments (generated using random initial concentration perturbations), which form a dataset comprising $N = \tilde N T/\Delta t$ data points.

As performance metric we use the area under the curve (AUC) $\mathcal{A}$ of the Receiver Operating Characteristic, indicating the probability that the $\hat\beta$ of a randomly chosen boundary species is smaller than that for a random interior species~\cite{harsh_see_nodate}; $\mathcal{A}=1$ indicates perfect separability, while $\mathcal{A}=0.5$ represents random guessing. \Cref{fig3}(a,b) shows $\mathcal{A}$ as a function of $\tilde N$ and $\Delta t$, for two different $T$ values. We find that $T$ is not crucial for performance, provided it exceeds the typical system relaxation time scale ($\approx 0.1$ for the parameters used~\cite{harsh_see_nodate}). The best discrimination is achieved when $\tilde N \geq 4 $ and $\Delta t \leq 10^{-2}$, which is the shortest relaxation time in the system. Along curves of constant number of data points $N$, performance is generally improved by increasing data diversity (larger $\tilde{N}$) at coarser resolution (larger $\Delta t$). Further analysis shows that this arises because, as long as %
reasonable temporal resolution is maintained, the regression for the boundary species gets {\em worse} ($\hat \beta_{\text{boundary}}$ decreases) with increasing $\tilde N$, while the regression for the interior species and its large $\hat \beta_{\text{interior}}$ change negligibly~\cite{harsh_see_nodate}.

\textbf{Time scales}. So far we have used parameters~\cite{harsh_see_nodate} with comparable  subnetwork and bulk reaction rates. Varying these systematically we find that if the bulk is comparatively slow one simply needs to observe longer trajectories, possibly at larger $\Delta t$, to capture all relaxation time scales 
(fig.~F4 in~\cite{harsh_see_nodate}). In the case of fast bulk reactions memory functions decay rapidly but surprisingly the boundary units remain detectable even with the larger $\Delta t$ appropriate for the subnetwork dynamics~\cite{harsh_see_nodate}.

\textbf{Conservation laws}. If the reaction network has conservation laws involving subnetwork species, the dynamics starting from a given initial condition may only explore an (affine) linear subspace of all $\bm{x}$, making the basis functions degenerate.
However, our method remains effective if the $\tilde{N}$ time series exhibit sufficiently diverse values of the conserved concentration combinations~\cite{harsh_see_nodate}.

\textbf{Enzymatic reactions}. Enzymatic reactions as described by Michaelis-Menten (MM) kinetics~\cite{MichaelisMenten2007} differ from the simple mass action reactions considered thus far, with a single enzymatic conversion $X_s \to X_p$ of substrate to product described by
\begin{equation}
\label{eq:MM_kinetics}
    \partial_t x_p = \frac{V_m x_s}{K_m + x_s}
\end{equation}
where $V_m$ and $K_m$ are the MM constants. Treating the enzymes as fast unobserved species, this dynamics can be effectively reduced~\cite{rubin_michaelis-menten_2016} by linearizing the numerator around the steady state value $y_s$ of $x_s$, giving  $\partial_t x_p = \hat \lambda_{sp} \delta x_s$ with
\begin{equation}
    \hat \lambda_{sp} = \frac{V_m}{K_m(1+y_s/K_m)^2}\left(1- \frac{\delta x_s/y_s}{K_m/y_s + 1} \right)
    \label{eq:linear_approx_MM}
\end{equation}
and $\delta x_s=x_s-y_s$. The effective reaction flux $\partial_t x_p$ is then again quadratic in concentrations and our method is applicable.

\textbf{EGFR network}. We next consider a biological example combining the above features, i.e.\ large spread of time scales, conservation laws and enzyme kinetics. The Epidermal Growth Factor Receptor (EGFR) network is a signalling pathway with downstream components that regulate cell growth, survival, proliferation etc.~\cite{oda_comprehensive_2005,steven_wiley_computational_2003}. Using a parametrization from kinetic studies~\cite{kholodenko_quantification_1999}, we simulate noisy time trajectories, starting from random perturbations around the steady state. Out of $P=29$ species (enzymes and their complexes are included implicitly via MM kinetics,~\cref{eq:MM_kinetics}), we take a biologically motivated subnetwork of $d=15$ species~\cite{rubin_memory_2014} with  four boundary species called SOS, GS, Grb and RP. 

In the noise free case, our method for identifying boundary species continues to work well~\cite{harsh_see_nodate}. Given the larger $P$ we need a larger number of independent experiments, $\tilde N \geq 60$, with $\Delta t \leq 0.05$s to resolve the typical relaxation dynamics. 
Interestingly, $T \geq 5$s suffices despite the long relaxation times of $\approx 100$s.%

\begin{figure}
\includegraphics[width=1.0\linewidth]{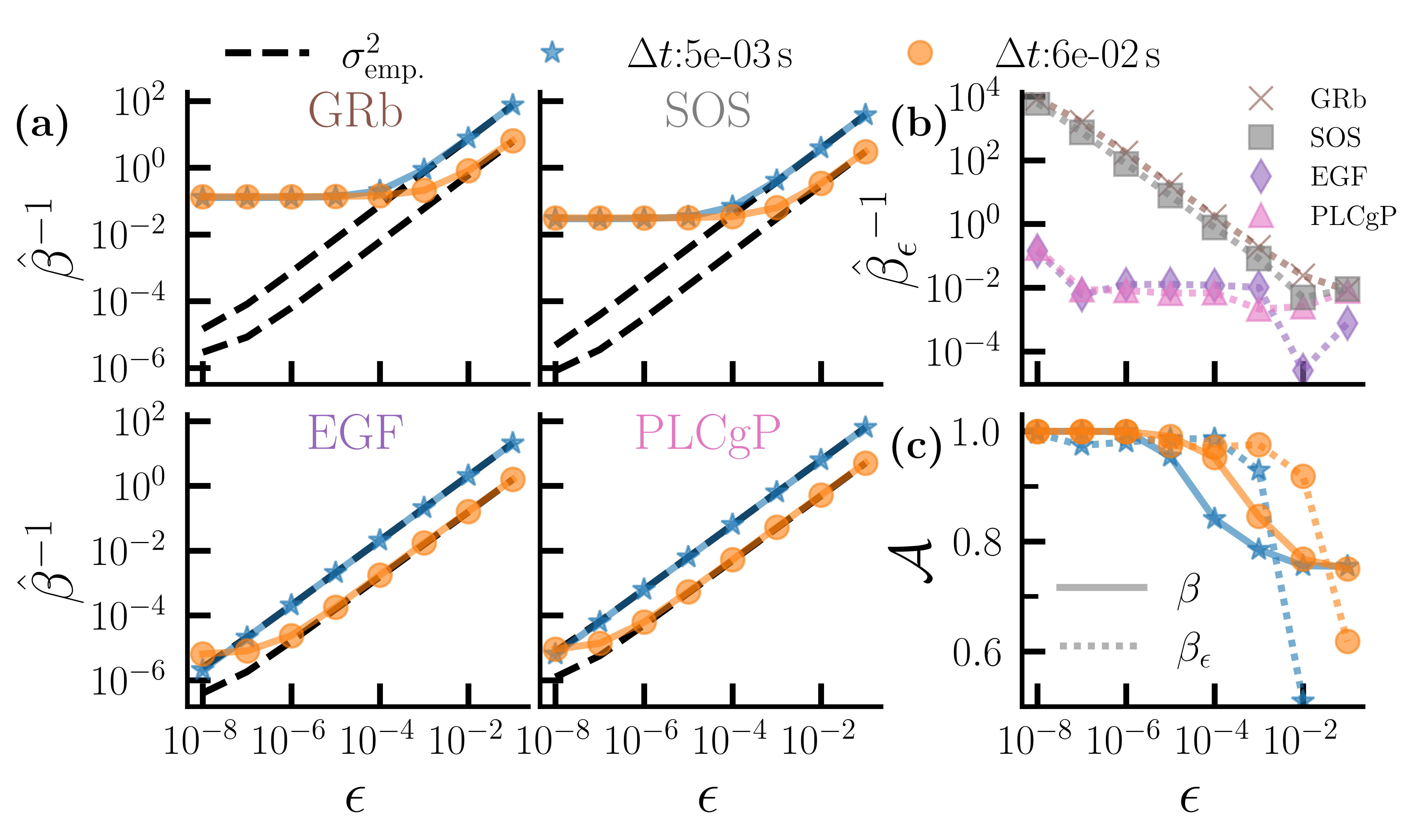}%
\caption{\label{fig4} (a) Effective noise variance $\hat \beta^{-1}$ versus %
noise level $\epsilon$ for two different $\Delta t$ (blue and orange) for two boundary species (top: GRb, SOS) and two interior species (bottom: EGF, PLCgP) of the EGFR network. Black dashed line: steady state empirical variance $\sigma^2_{\text{emp}}$. 
(b) Relative effective noise variance $\hat \beta^{-1}_\epsilon$ for the four species for $\Delta t = 5\times10^{-3}$s, demonstrating that discrimination is possible up to noise levels $\epsilon < 10^{-3}$. (c) AUC $\mathcal{A}$ against $\epsilon$ for two different $\Delta t$. Using $\beta_\epsilon$ (dashed) rather than $\beta$ extends the applicability of the method to higher noise levels.}
\end{figure}

\textbf{Effect of noise}. With stochastic noise we expect boundary detection to continue to work as long as the effective noise on the boundary species from memory and random force is larger than the stochastic effects. For a typical cell of volume $V =(20\;\mu \mathrm{m})^3$, $\epsilon = 1/V$ is of order $10^{-4}$\,nM. 
To quantify the empirical noise we measure the variance $\sigma^2_{\text{emp}}$ of $\partial_t x_i$ in steady state for each molecular species. \Cref{fig4}(a) shows that as we increase the noise level $\epsilon$ in the EGFR network, the effective noise variance $\hat \beta^{-1}$ for boundary species is initially nearly constant as expected, but for large $\epsilon$ becomes dominated by $\sigma^2_{\text{emp}}$. We therefore propose to use for noisy systems the relative (excess) effective noise variance
\begin{equation}
    \hat \beta_\epsilon^{-1} = \frac{\hat \beta^{-1} - \sigma^2_{\rm{emp}}}{\sigma^2_{\rm{emp}}}
    \label{eq:beta_eps_def}
\end{equation}
to distinguish interior and boundary species as shown in~\cref{fig4}(b). This gives excellent discrimination performance~(\cref{fig4}(c)) up to larger $\epsilon$ than using the bare $\hat\beta$. Note that for stochastic dynamics, performance is better for {\em larger} (but not exceeding dynamical timescales) $\Delta t$ as it reduces noise in finite difference estimates of $\partial_t x_i$. 

\begin{figure}[ht]
\includegraphics[width=0.95\linewidth]{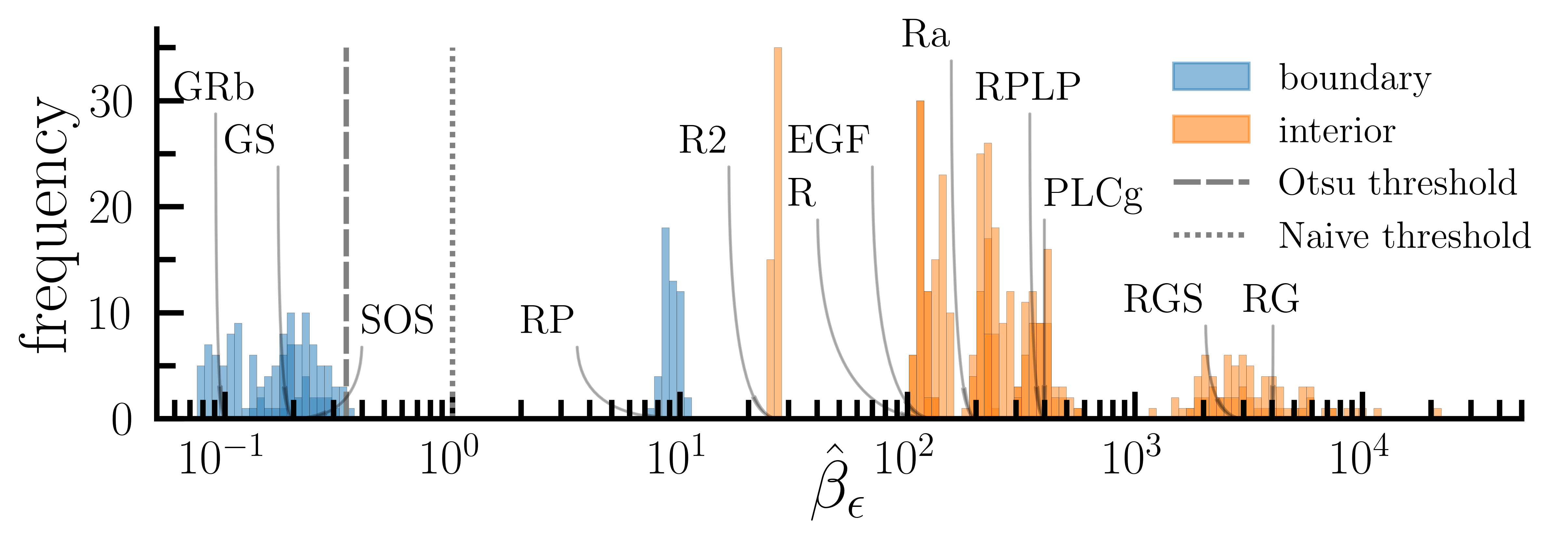}%
\caption{\label{fig5} Histogram of $\hat \beta_\epsilon$ across 50 independent data sets at noise level $\epsilon=10^{-5}$ for all observed species of the EGFR network (blue: boundary species, orange: interior species); the mean $\hat\beta_\epsilon$ for some species are labelled explicitly. Both the naive (vertical dotted grey line at unity) and Otsu thresholds (vertical dashed grey line) correctly classify GRb, GS and SOS.}
\end{figure}

Beyond the AUC, for practical classification one requires a threshold on $\hat \beta$ or $\hat\beta_\epsilon$. The naive value is unity: from~\cref{eq:beta_eps_def} we have effective noise significantly in excess of stochastic noise, indicating a boundary species, when $\hat\beta_\epsilon^{-1}>1$. For more precise automatic threshold detection one could use e.g.\ the Otsu algorithm~\cite{otsu_threshold_1979}, which minimizes the weighted sum of the two intra-class variances. In~\cref{fig5} we plot the histogram of $\log \hat \beta_\epsilon$ for EGFR. Both the naive and the Otsu thresholds classify GRb, GS and SOS correctly as boundary species, but miss RP since it undergoes extremely slow bulk reactions as compared to its subnetwork reactions (by six orders of magnitude); the value of $T$ used is thus too small to discern appreciable memory effects.

Summarizing, we have proposed and tested a method for identifying boundary species in partially observed reaction networks that does not rely on additional {\em a priori} information about unknown parts of a network, In addition to making existing network topology reconstruction techniques~\cite{willis_inference_2016,burnham_inference_2008,langary_inference_2019,bonvin_target_1990} more robust, the approach can form the basis for systematic expansion of the network, or for targeting efforts to reconstruct memory functions~\cite{meyer_non-markovian_2020, jung_iterative_2017, kowalik_memory-kernel_2019,meyer_numerical_2021,doerries_correlation_2021} at the boundary species  and thus extract information about the unobserved network. The broad framework of our method also suggests interesting generalizations worth exploring, e.g.\ extensions to more strongly nonlinear models as used in~\cite{ji_autonomous_2021}, to non-parametric Gaussian process regression~\cite{rasmussen_gaussian_2006}, or to Relevance Vector Machines~\cite{bishop_pattern_2006} that could fit networks with widely varying 
rate constants using different weight priors for each basis function. 

\bibliography{paper_main.bib}

\end{document}


\maketitle	

\section{Reaction network dynamics and memory functions
} \label{sec:MAK}
	
We outline here the general type of reaction network we consider and the structure of the memory effects arising from partial observations, where only a subnetwork is tracked. The formalism is also illustrated using the simple five species network from the main text. 

We study chemical reaction networks defined as systems of chemical reactions 
	\begin{equation}
	    \sum_{i=1}^P s_{ir} X_i \xrightarrow[\text{}]{k_r} \sum_{i=1}^P {s}'_{ir} X_i %
	\end{equation}
We focus on the standard case where the reactions are at most bimolecular ($\sum_i s_{ir}\leq 2$, $\sum_i s'_{ir}\leq 2$); higher order reactions can be represented as sequences of such bimolecular ones. We assume that the reaction flux (propensity function) for each reaction follows mass-action kinetics (MAK):
	\begin{equation}
	    f_r = k_r \prod_{i=1}^{P} x_i^{s_{ir}}
	\end{equation}
where $x_i$ is the concentration of species $X_i$ and $k_r$ has appropriate dimensions. The entries $S_{ir}=s'_{ir}-s_{ir}$ of the stochiometric matrix $\bm{S}$ give the net change in molecule numbers for each reaction.~\Cref{table:different_reactions} collects the types of reactions that can occur and gives the reaction fluxes and stoichiometries. 

For large enough reaction volumes one can work with continuous concentration variables and neglect noise arising from the discrete nature of copy numbers~\cite{harsh_accurate_2023}. Such noise can be potentially accounted for using different techniques,~e.g. by employing memory~\cite{harsh_accurate_2023,fromberg_generalized_2021}.
Using the stoichiometric matrix and the vector $\bm{f}$ of reaction fluxes the resulting dynamics is compactly represented as
	\begin{equation}
	    \partial_t \bm x^* = \bm S \bm f 
	    \label{eq:compact_time_evolution}
	\end{equation}
 
To account for fluctuations around the average concentrations to leading order in $\epsilon = V^{-1}$ we also consider reaction kinetics described by the Chemical Langevin Equation~\cite{schnoerr_approximation_2017}:%
	\begin{equation}
	    \partial_t \bm x^* = \bm S \bm f + \bm \eta
	\end{equation}
with the noise correlation given by $\langle \boldsymbol{\eta}(t) \boldsymbol{\eta}^{\rm T}(t') \rangle = \epsilon \, \bm S\, \text{diag} (\bm f)\,\bm S^{\rm T} \delta(t - t')$.

	\begin{table}[ht]
	\centering
	\begin{tabular}{| c|c|c|c|}
		\hline
		Reaction & Type & Fluxes, $f_i$ & Stoichiometry%
		\\
		\hline
		$X_j + X_l \xrightarrow{k_{jl,i}} X_i $ & Complex formation & $ k_{jl,i}x_j x_l $ & $S_j=S_k=-1$, $S_i=+1$ \\
        \hline
        $X_j + X_l \xleftarrow{k_{i,jl}} X_i $ & Complex dissociation & $ k_{i,jl} x_i $ & $S_j=S_k=+1$, $S_i=-1$ \\
        \hline
        $X_i \xrightarrow{\lambda_{i,j}} X_j $  & Unary conversion & $ \lambda_{i,j}x_i $ & $S_i=-1$, $S_j=+1$ \\
		\hline
		$X_i \xrightarrow{\omega_{i}} \emptyset $ & Destruction & $ \omega_i x_i $ & $S_i=-1$\\
        \hline
  		$X_i \xleftarrow{\kappa_{i}} \emptyset $ & Creation & $ \kappa_i $ & $S_i=+1$\\
		\hline
	\end{tabular}
	\caption{Possible reactions in a network with at most bi-molecular reactions, with the corresponding reaction fluxes and stoichiometries.}
	\label{table:different_reactions}
	\end{table}
    
    \begin{table}[t]
        \centering
        \begin{tabular}{rrrr}
        ${k}_{12, 3}$ & ${k}_{3,12}$ & ${k}_{14,5}$ & ${k}_{5,14}$ \\ \hline
        \rowcolor[HTML]{EFEFEF} 
        10                   & 1.5                 & 5.8                 & 10                 
        \end{tabular}
        \caption{%
        Dimensionless rate constants for the five species network. All simulations for this network are carried out with these rate constants unless stated otherwise.}
        \label{tab:appendix:simple_network}
    \end{table}
 
	\begin{figure}
    \centering
    \includegraphics[width=.75\linewidth]{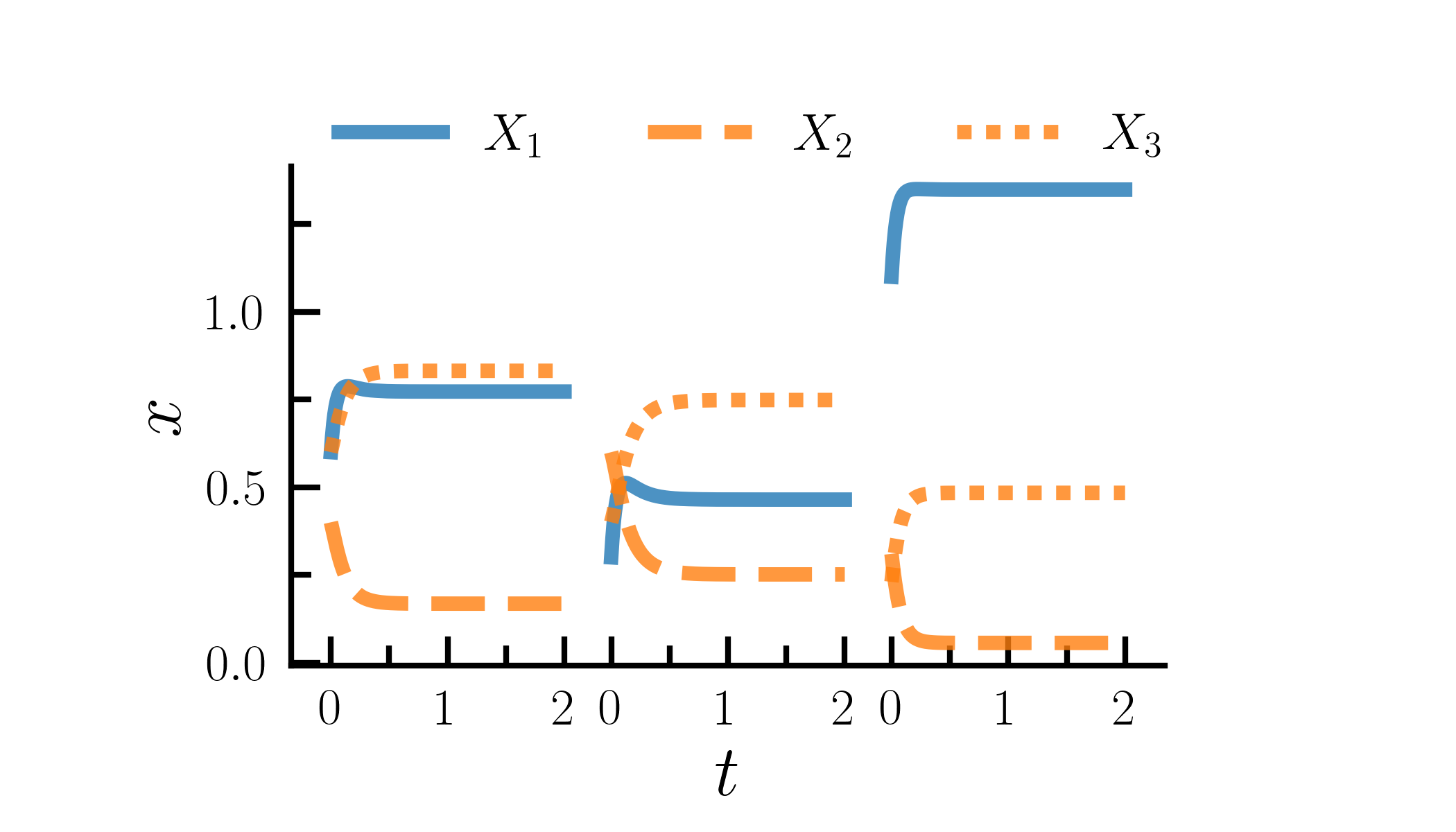}
    \caption{Time traces for the sub-network species in the simple network model with the parameters from~\cref{tab:appendix:simple_network} with $T = 2,\ \Delta t = 0.01$. Three different experiments ($\tilde N = 3$) are plotted here, starting from random initial conditions. $X_1$ is the boundary species while $X_2$ and $X_3$ are the interior species.}
    \label{fig:time_traces_simple_network}
\end{figure}
	
For the simple five species network from the main text the (noise free) dynamical equations are
        \begin{equation}
        \begin{split}
        \partial_t x_1 &= -k_{12,3} x_2 x_1 + k_{3,12} x_3 - k_{14,5} x_4 x_1 + k_{5,14} x_5\\
        \partial_t x_2 &= -k_{12,3} x_2 x_1 + k_{3,12} x_3\\
        \partial_t x_3 &= k_{12,3} x_2 x_1 - k_{3,12} x_3\\
        \partial_t x_4 &= - k_{14,5} x_4 x_1 + k_{5,14} x_5 \\
        \partial_t x_5 &= k_{14,5} x_4 x_1 - k_{5,14} x_5
        \label{eq:theory:simple_network_mak}
        \end{split}
        \end{equation}
The rate constants are given in~\cref{tab:appendix:simple_network} and some sample time traces are shown in~\cref{fig:time_traces_simple_network}.
        
We now turn our attention to a subnetwork consisting of $d$ out of the total $P$ molecular species that are observed, with their concentrations being given by $\bm x$. The dynamics of the subnetwork using the linear and quadratic basis functions $\bm \phi(\bm x)$ as explained in the main text can be written explicitly as
	\begin{equation}
        \label{eq:projection_non_linear}
        \begin{split}
        \partial_t x_i (t) = W_{i} &+ \sum_{j=1}^d x_j(t) W_{j,i} + \sum_{j,k,j\leq k}^d x_j(t) x_k(t) W_{jk,i}  + \int_0^t dt' \left( \sum_{j=1}^d x_j(t') \Gamma_{ji} (t-t') \right. \\
        &\left. +\sum_{j,k,j\leq k}^d x_j(t') x_k(t') \Gamma_{jk,i}(t-t') \right) + r_i (t) 
        \end{split}
    \end{equation}
	The derivation within the projection formalism leading to this as well as the resulting expressions for the local-in-time coefficients $\bm W$, the memory functions $\bm \Gamma$ and the random force $\bm r$ can be found in~\cite{rubin_memory_2014,herrera-delgado_memory_2018}. The key feature of these expressions is that the memory and random force terms arise only for the boundary species, while for the subnetwork species the first three terms completely capture the dynamics.
We illustrate this projection formalism for the five species network in the main text. For this it is convenient to switch from the variables $\bm x$ to the deviations from the steady state $\bm y$, i.e~$\bm \delta x = \bm x - \bm y$. The dynamics of the boundary species $X_1$ then takes the form~\cite{herrera-delgado_memory_2018,rubin_memory_2014}
	\begin{align}
    \begin{split}
        \partial_t \delta x_1(t)  = &-\delta x_1(t)(k_{12, 3} y_2 + k_{14, 5} y_4) -\delta x_2(t) k_{12, 3} y_1  + \delta x_3(t) k_{3,12}\\
        &+\int_0^t dt'\delta x_1(t^\prime) \Gamma_{1,1}(t-t^\prime) + r_1(t)
    \end{split}
    \label{eq:appendix:projected_equation_simple_network}
    \end{align}
	where for simplicity we have only kept linear terms in $\delta\bm{x}$.
	The first three terms are the reaction terms for $\delta x_1$ (that only include the first three species). The memory function originates from the projection onto the subnetwork $X_1,X_2,X_3$ and is given by
	\begin{equation}
    \Gamma_{1,1}(t-t^\prime) = k_{14,5} y_4 (k_{14,5} y_1 + k_{5,14}) e^{-(k_{14,5}y_1 + k_{5,14})(t-t^\prime)}
    \label{eq:appendix:M_linearized_dynamics}
    \end{equation}
    The random force reflecting the dependence on the initial concentration deviation from steady state of $X_4$ and $X_5$ is
    \begin{equation}
        r_1(t)= (\delta x_5(0) k_{5,14} - \delta x_4(0)) e^{-(k_{14,5} y_1 + k_{5,14}) t}
    \label{eq:appendix:random_force_simple_network}
    \end{equation}
    As $X_2$ and $X_3$ are internal species (they only interact with other subnetwork species), their memory functions and random forces vanish.

    \begin{figure}
      \centering
      \includegraphics[width=.95\linewidth]{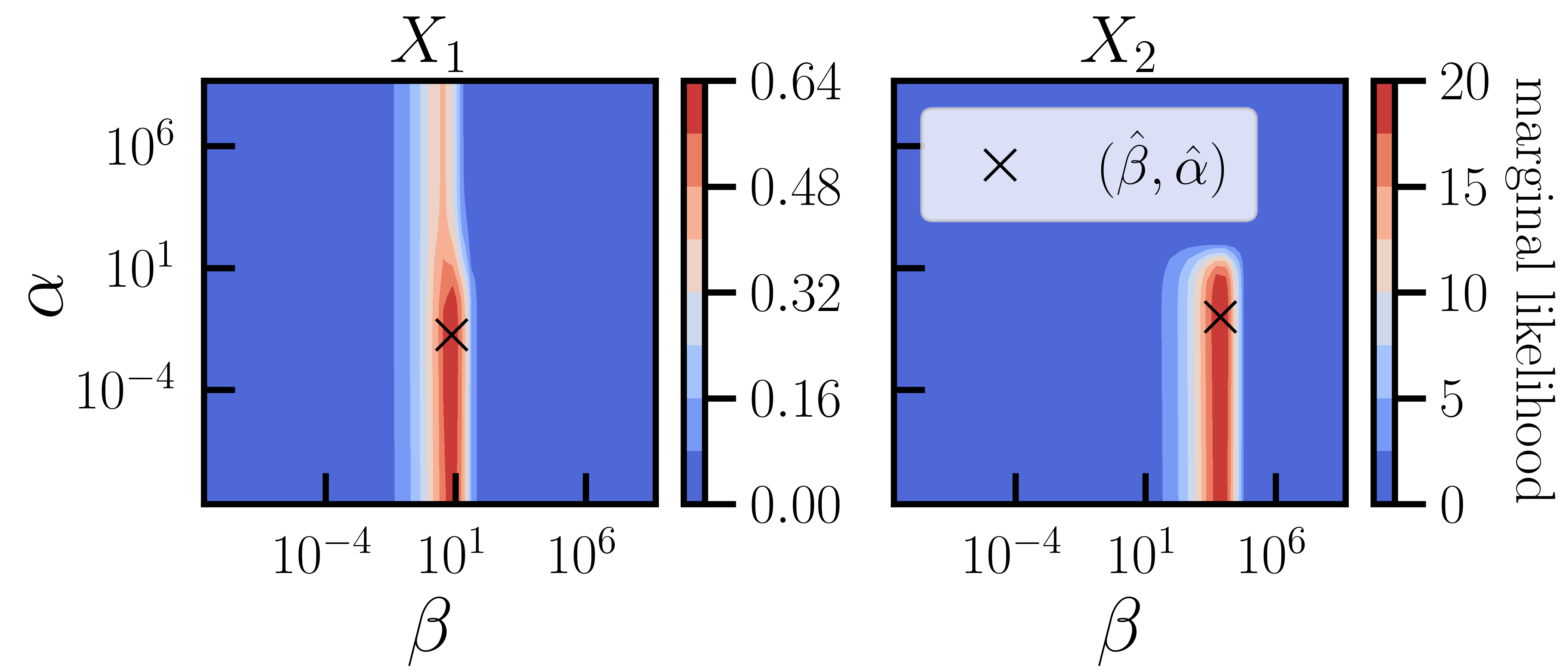}
      \caption{Marginal likelihood landscape, $p(\bm \tau_i | \alpha_i,\beta_i, \bm \Xi)$, for $X_1$(left) and $X_2$(right) of the simple five species network. The maxima determining the optimal hyper-parameters are at $\hat \alpha_1 = 0.02$, $\hat \beta_1 = 7.12 $, $\hat \alpha_2 = 0.10$, $\hat \beta_2 = 7631.36$. These hyper-parameters are used in Fig.~1 of the main paper. The network parameters are taken from~\cref{tab:appendix:simple_network} with simulation parameters being $\tilde N = 10,\ T = 2,\ \Delta t = 0.01$.}
      \label{fig:likelihood landscape}
  \end{figure}

\section{Determination of hyperparameters}
\label{sec:likelihood_optimization}

\begin{figure}[t]
      \centering
      \includegraphics[width=.80\linewidth] {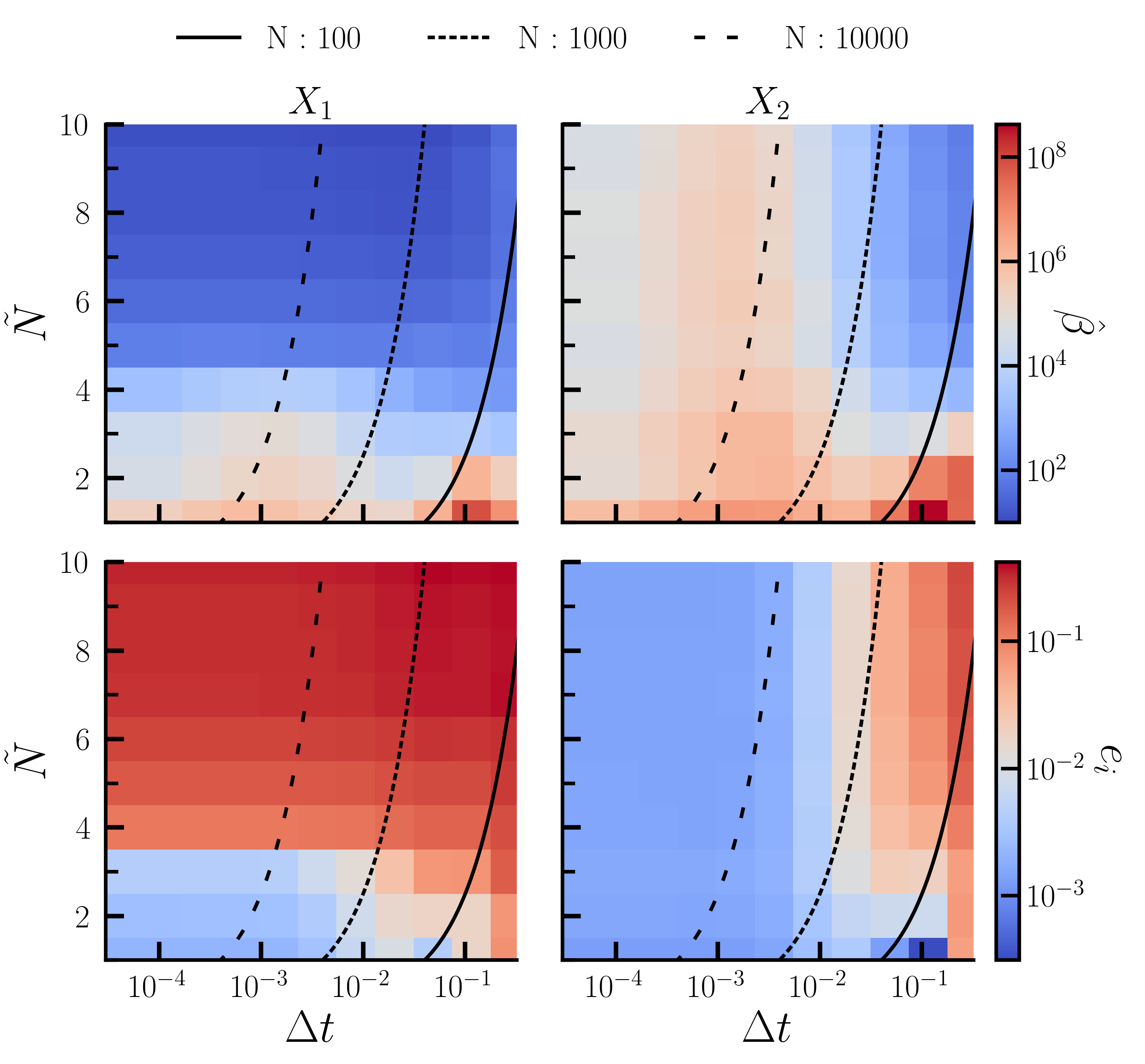}
        \caption{(Top) The estimated $\hat \beta_1$ (left) and $\hat \beta_2$ (right) as we change number of experiments, $\tilde N$, and time step, $\Delta t$ with fixed $T = 4$; results are averaged over 10 independent data sets for each ($\tilde N,\ \Delta t$) pair.
        (Bottom) The regression error, $e_i$ as defined in \cref{eq:regression_error}, for the same parameters as in the top and for the two species $X_1$ (left) and $X_2$ (right). The number of training data points $N$ is constant across each of the three black lines. Our method works in the parameter regime approximately bounded by $\tilde N > 4$ and $\Delta t \leq 10^{-2}$, where $\hat\beta_2 \gg \hat\beta_1$. This figure is plotted from the same data as used in Fig.~2 of the main text, with network parameters taken from~\cref{tab:appendix:simple_network}.}%
      \label{fig:beta_values_par}
  \end{figure}
  
In this section we describe how to obtain the optimal (hyper-)priors $\hat\alpha\text{ and } \hat \beta$ using the Bayesian model selection approach outlined in the main text. 

The weights $\bm{w}_i$, as stated in the main text, are taken to have a zero mean Gaussian prior $p(\bm{w}_i|\alpha_i)$ with covariance matrix $\alpha_i \mathbb{I}$. The likelihood of the observations $\bm \tau_i$ as in eq.~(2) of the main text, given the prior on the weights and the Gaussian observation noise of variance $1/\beta_i$ is then given by
\begin{equation}
p(\bm{\tau}_i | \bm{w}_i, \bm\Xi, \beta_i) = \prod_n \mathcal{N}(\tau_i^{(n)} | \bm{w}_i^{\rm T}\bm\phi(\bm{x}^{(n)}),1/\beta_i)
\end{equation}
where $\mathcal{N}(\cdot | \mu,\sigma^2)$ indicates a Gaussian distribution with mean $\mu$ and variance $\sigma^2$. The joint probability of observations and weights is then
\begin{equation}
p(\bm{\tau}_i,\bm{w}_i | \bm{\Xi},\alpha_i,\beta_i) = p(\bm{\tau}_i| \bm{w}_i, \bm{\Xi},\beta_i)
p(\bm{w}_i | \alpha_i) 
\end{equation}
Integrating over the weights $\bm w_i$ gives the marginal likelihood $p(\bm{\tau}_i | \alpha_i,\beta_i,\bm{\Xi})$ and as discussed in the main text, the posterior distribution of the hyperparameters $\alpha_i$, $\beta_i$ is proportional to this marginal likelihood if one uses Bayes' theorem with a flat hyperprior. 
Explicitly the (log) marginal likelihood is
  \begin{equation} 
    \begin{split}   
      \label{eq:theory:log_marginal_likelihood} 
      \ln p(\bm \tau_i | \alpha_i, \beta_i, \bm{\Xi}) &= \frac{M}{2} \ln \alpha_i + \frac{N}{2} \ln \beta_i - \frac{\alpha_i}{2} \bm m_i^\text{T} \bm m_i -\frac{\beta_i}{2} || \bm \tau_i -  \bm \Phi \bm m_i ||^2 -\frac{1}{2}\ln |\bm \Sigma_i^{-1}| - \frac{N}{2} \ln(2\pi)
    \end{split} 
  \end{equation}
where the design matrix $\bm\Phi$ as defined in the main text 
has elements $\Phi_{nm} = \phi_m(\bm x^{(n)})$, $M = d(d+3)/2+1$ is the total number of basis functions,  
$\bm m_i = \beta_i \bm \Sigma_i \bm \Phi^\text{T} \bm \tau_i$ and $\bm \Sigma^{-1}_i = \alpha_i \mathbb{I} + \beta_i \bm \Phi^\text{T} \bm \Phi $. 
The optimal hyperparameters $\hat \alpha_i, \hat \beta_i$ are found by maximizing this quantity. This can be done by gradient ascent using the explicit expressions
\begin{equation}
    \label{eq:gradient_log_likleihood}
    \begin{split}
    \frac{\partial \ln p}{\partial \alpha_i} &= \frac{M}{2\alpha_i} - \frac{1}{2}\bm m_i^\text{T} \bm m_i - \frac{1}{2} \text{Tr} (\bm \Sigma_i^{-1})\\
    \frac{\partial \ln p}{\partial \beta_i} &= \frac{N}{2\beta_i} - \frac{1}{2}||\bm \tau_i -  \bm \Phi \bm m_i ||^2 -\frac{1}{2} \text{Tr}(\bm \Sigma_i^{-1}\bm \Phi^\text{T} \bm \Phi)
    \end{split}
\end{equation}
In practice we use the more efficient \texttt{linear\_model.BayesianRidge}~\cite{tipping_sparse_2001,mackay_bayesian_1992} method of the \texttt{Python} \\ \texttt{Scikit-learn}~\cite{pedregosa_scikit-learn_2011} library.

In~\cref{fig:likelihood landscape} we plot as an illustration the marginal likelihood $p(\bm \tau_i | \alpha_i,\beta_i, \bm \Xi )$ as a function of $\alpha_i$ and $\beta_i$ for $X_1$ and $X_2$ for the simple five species network from the main text.
Crosses indicating the positions $(\hat \alpha_i, \hat \beta_i)$ of the maxima are also shown, which are used in Fig.~1 of the main paper. The likelihood is highly peaked in the $\beta$ direction, but is relatively broader in the $\alpha$ direction. Thus our estimate of the effective noise is much more precise than the estimate of the prior variance of the weights, and it is the former %
that we are primarily interested in.

After determining $\hat\alpha_i$ and $\hat\beta_i$, the mean (or most likely) posterior weights can be calculated as
\begin{equation}
    \hat{\bm{w}}_i 
= \hat \beta_i \bm{\hat \Sigma}_i \bm \Phi^{\rm T} \bm \tau_i 
\end{equation}
with $\bm{\hat \Sigma}^{-1}_i = \hat \alpha_i \mathbb{I} + \hat \beta_i \bm \Phi^{\rm T} \bm \Phi$~\cite{bishop_pattern_2006}. 
The predictive distribution of the target $\tau_i$ for a new set of subnetwork concentrations $\tilde{\bm x}$ is Gaussian, with mean and variance given by
\begin{equation}
    \hat \tau_i(\tilde{\bm x}) = \hat{\bm w}_i^{\rm T} \bm \phi(\tilde{\bm x}), \quad \quad \hat \sigma^2_i(\tilde{ \bm x}) = \hat \beta^{-1}_i \mathbb{I} + \bm \phi^{\rm T}({\tilde{\bm x}}) \bm{\hat \Sigma}_i \bm \phi({\tilde{\bm x}})
\end{equation}

Except for very small training data sets the marginal likelihood and hence also the posterior $p(\alpha_i,\beta_i | \bm\tau_i, \bm\Xi)$ %
is sufficiently peaked to be well approximated by a Gaussian around $\hat \alpha_i, \hat \beta_i$. The covariance of this Gaussian can be calculated from the Hessian of the log-likelihood, which is %
\begin{equation}
    \label{eq:hessian_lanscape}
    \begin{split}
    \frac{\partial^2 \ln p}{\partial \alpha_i^2} &= -\frac{M}{2\alpha_i^2} + \bm m_i^\text{T}\bm \Sigma_i^{-1} \bm m_i + \frac{1}{2} \text{Tr}(\bm \Sigma_i^{-2})\\
    \frac{\partial^2 \ln p}{\partial \beta_i^2} &= -\frac{N}{2\beta_i^2} + \frac{1}{2} \text{Tr}((\bm \Sigma_i^{-1}\bm \Phi^\text{T} \bm \Phi)^2) - (\bm \tau_i -  \bm \Phi \bm m_i)^\text{T} \left( \bm \Phi \bm \Sigma_i^{-1} \bm \Phi^\text{T} \bm \Phi \bm m_i - \beta_i^{-1} \bm \Phi \bm m_i \right)\\
    \frac{\partial^2 \ln p}{\partial \alpha_i \partial \beta_i} &= -(\bm \tau_i -  \bm \Phi \bm m_i)^\text{T} ( \bm \Phi \bm \Sigma_i^{-1} \bm m_i) + \frac{1}{2} \text{Tr}(\bm \Sigma_i^{-2} \bm \Phi^\text{T} \bm \Phi)
    \end{split}
\end{equation}
The covariance of the Gaussian approximation to the posterior is then the Fisher information matrix
\begin{equation}
    \label{eq:hessian matrix}
    \bm{F}_i = -
    \begin{pmatrix}
        \frac{\partial^2 \ln p}{\partial \alpha_i^2} & \frac{\partial^2 \ln p}{\partial \alpha_i \partial \beta_i}\\
        \frac{\partial^2 \ln p}{\partial \alpha_i \partial \beta_i}& \frac{\partial^2 \ln p_i}{\partial \beta_i^2}
    \end{pmatrix}^{-1}%
\end{equation}
 evaluated at the optimal hyperparameters $\hat \alpha_i, \hat \beta_i$. 
The uncertainties in the hyperparameters, $\delta \hat \alpha_i$, and $\delta \hat \beta_i$ are given by the diagonal elements
\begin{equation}
\label{eq:fisher_info_sigma}
    \delta \hat \alpha_i = \sqrt{(\bm{F}_i)_{11}}, \quad \delta \hat \beta_i = \sqrt{(\bm{F}_i)_{22}}
\end{equation}

We use these uncertainties in our calculation of AUCs: we pick a random interior species (e.g.\ $X_2$), sample $\beta_2$ from a Gaussian distribution with mean $\hat\beta_2$ and standard deviation $\delta\hat\beta_2$, and then do the same for the boundary species $X_1$. The AUC $\mathcal{A}$ as defined in the main text is then the probability that $\beta_2>\beta_1$. For the Gaussian distributions at hand it can be expressed in closed form once the $\hat\beta_i$ and $\delta\hat\beta_i$ are known. 

We illustrate the optimal hyperparameter values in~\cref{fig:beta_values_par}~(top), where we plot the $ \hat \beta$ for $X_1$ and $X_2$ as we vary $\tilde N$ and $\Delta t$ at fixed $T = 4$;  network parameters are as in from~\cref{tab:appendix:simple_network} and results are averaged over 10 independent data sets for each $\tilde N$, and $\Delta t$ pair.
We see here the effects mentioned in the main text: for the boundary species, $\hat \beta$ is predominantly affected by $\hat N$; it decreases 
for large $\tilde N$ because the regression gets worse and the effective noise due to the memory terms increase. For the interior species, on the other hand, given small enough time steps $\Delta t < 10^{-2}$, i.e.\ sufficient time resolution, the regression performs well and results in a large $\hat \beta$ that is essentially independent of $\tilde N$.

The same trends can be seen in~\cref{fig:beta_values_par}~(bottom) %
where we plot the time averaged error between the observed time derivative and its regression estimate, defined as
\begin{equation}
    e_i = \frac{1}{N}\sum_{n=1}^{N} |\tau_i^{(n)} - \hat \tau_i^{(n)}|
    \label{eq:regression_error}
\end{equation} 
which is further averaged over the ten independent data sets. Again the regression quality is affected predominantly by $\tilde N$ for the boundary species and by $\Delta t$ for the interior species. The error increases with $\tilde N$ for the boundary species, as the regression is less and less able to present the memory terms in the form of effective reactions. For the interior species, by contrast, the already small regression error decreases with $\tilde{N}$ as more and more data leads to more precise estimates.

\section{Timescale effects}
\label{sec:results:bulk_rate_scaling}

We give details here of the performance of our method in the regime where the reactions of a boundary species with the bulk species are significantly different from those with other subnetwork species. The former govern the rate with which the memory functions decay, so if they are fast then the memory will decay rapidly on the scale of the subnetwork dynamics. This might lead one to expect that a memory-based boundary detection approach would fail. However, we will demonstrate that this intuition is too simple and that the approach does in fact still work.

We study this scenario for the simple network with five species. We assume original reaction rate constants of $O(1)$ and then scale the rates of the interaction of the subnetwork with the bulk by a factor of $\gamma$ such that
\begin{equation}
\label{eq:bulk-rate-scaling}
    k_{14,5} \rightarrow \gamma k_{14,5}, \; k_{5,14} \rightarrow \gamma k_{5,14}
\end{equation}
In the limit of large $\gamma$, the bulk relaxes at short time scales of order $\sim 1/\gamma$. 
Consider now the linear memory function from  in~\cref{eq:appendix:projected_equation_simple_network}. With the scaling by $\gamma$ included, the dynamics of $X_1$ becomes
\begin{equation}
    \begin{split}
        \partial_t \delta x_1(t)  &= -k_{12, 3}(y_2\delta x_1(t) + y_1 \delta x_2(t)) + k_{3,12} \delta x_3(t) - \gamma \delta x_1(t) k_{14, 5} y_4 \\&+  \gamma^2 \int_0^t dt'\delta x_1(t^\prime)  k_{14,5} y_4 (k_{14,5} y_1 + k_{5,14}) e^{-\gamma (k_{14,5} y_1 + k_{5,14})(t-t^\prime)} \\&+ \gamma (\delta x_5(0) k_{5,14} - \delta x_4(0)) e^{-\gamma (k_{14,5} y_1 + k_{5,14}) t} 
    \end{split}
    \label{eq:results:projected_equation_bulk_rate_scaling}
\end{equation}
In the limit of large $\gamma$, only short time differences $\Delta = (t-t')\sim 1/\gamma$ are important. A Taylor approximation of $\delta x_1(t')$ around time $t$ for small $\Delta$ gives, 
\begin{equation}
    \delta x_1 (t-(t-t')) = \delta x_1(t) - (t-t')  {\delta \dot x_1(t)} + \mathcal{}{O}((t-t')^2)
    \label{eq:results:taylor_expansion_delta_x_1}
\end{equation}
Evaluating the integral in eq. \refeq{eq:results:projected_equation_bulk_rate_scaling}, i.e.\ the memory term, in this approximation gives
\begin{equation}
    \begin{split}
    I(t) \approx k_{14,5} y_4 & \left [  \gamma \delta x_1(t)  - \frac{\delta \dot x_1(t)}{ k_{14,5}y_1 + k_{5,14}} \right. \\ 
    \hspace{2cm} & \left. + e^{-\gamma  (k_{14,5}y_1 + k_{5,14}) t} \left (- \gamma \delta x_1(t) + \gamma  \delta \dot x_1(t) t + \frac{\delta \dot x_1(t)}{ k_{14,5}y_1 + k_{5,14}}\right )\right ]
    \end{split}
    \label{eq:results:bulk_rate_scaling_integrated_memory}
\end{equation}
The final term decays exponentially with $t$ and so can be ignored in the large $\gamma$ limit for $t \gg 1/\gamma$. The same aplies to the random force term (last term) in~\cref{eq:results:projected_equation_bulk_rate_scaling}.
This yields for large $\gamma$
\begin{equation}
\partial_t \delta x_1 =
-k_{12, 3}(y_2\delta x_1 + y_1 \delta x_2) + k_{3,12} \delta x_3 - \frac{y_4}{y_1 + k_{5,14}/k_{14,5}}\delta \dot{x}_1
\label{eq:results:partial_t_delta_x1_limit_gamma}
\end{equation}
The first two terms on the right are just the linearized version of $ -k_{12,3} x_2 x_1 + k_{3,12} x_3 $. Reinstating the nonlinear form and solving for $\partial_t x_1 \equiv \delta\dot x_1$
then results in
\begin{equation}
    \partial_t x_1(t) \approx \frac{-k_{12,3} x_2 x_1 + k_{3,12} x_3}{1+\frac{y_4}{y_1 + k_{5,14}/k_{14,5}}}
    \label{eq:results:approx_partial_x_polynomial}
\end{equation}
The interaction of $X_1$ with the fast bulk that is in quasi-steady state with the subnetwork thus leads 
to dynamics that appears to be governed by effective rate constants
and can be represented by 
our basis functions. %
Crucially, however, 
different steady state values $y_1,y_4$ result in different effective rate constants so that there is no possibility of a consistent fit in a data set combining time series from multiple initial conditions. Our method will therefore still detect significant effective noise singling out the boundary species. 

\begin{figure} \centering
    \includegraphics[width=0.8\linewidth]{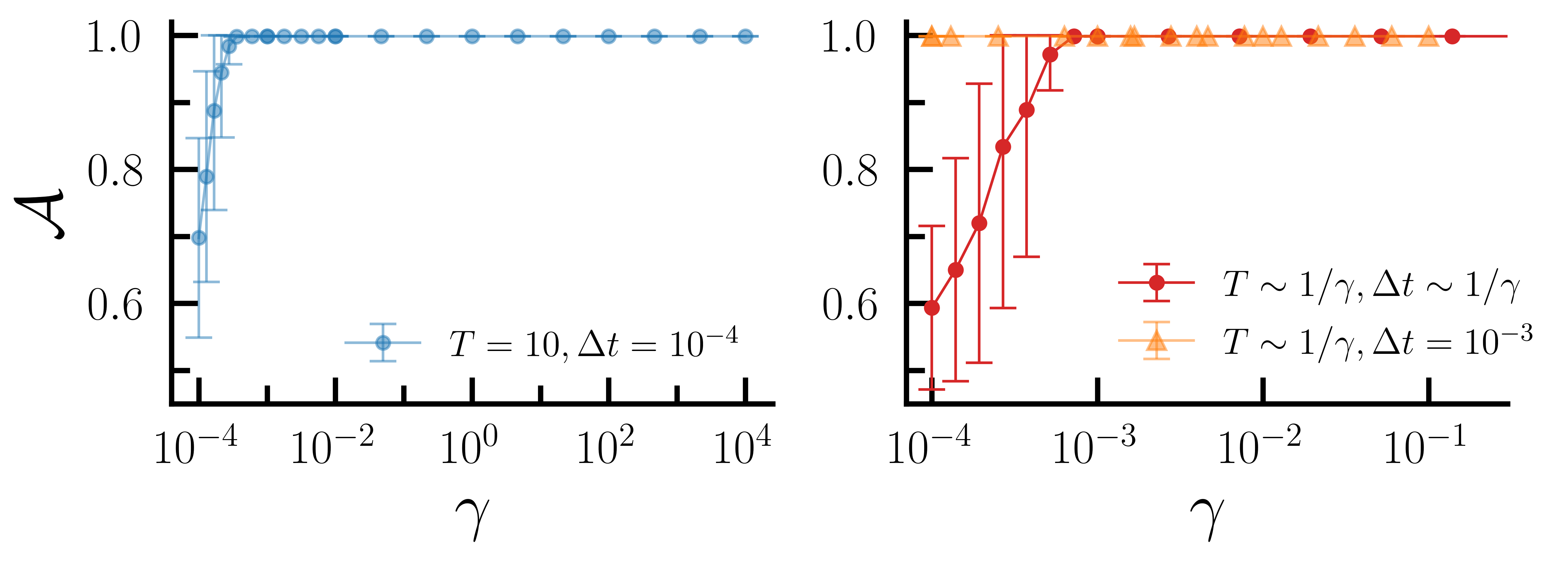}
    \caption{(Left) The AUC $\mathcal{A}$ as we scale the subnetwork-bulk reaction rates by $\gamma$. Results are averaged over five independent data sets each containing $\tilde N = 10 $ numerical experiments. The boundary detection method works reliably for large $\gamma$ as expected ($\mathcal{A} \approx 1 $), and in fact down to $\gamma \geq 10^{-3}$. (Right) In the small $\gamma$ range, keeping the other parameters same, we plot the AUC (red) if we scale $T = 0.64/\gamma$ and $\Delta t = 6.4\times 10^{-5}/\gamma$. We compare with the AUC (orange) for smaller scaled $T = 0.064/\gamma$ and fixed $\Delta t = 10^{-3}$.  The comparison shows that our method works reliably for small $\gamma$ if we can cover both the characteristic time of the bulk of $T=O(1/\gamma)$ {\em and} resolve %
    the subnetwork dynamics with a small enough $\Delta t$.    
    }
    \label{fig:results:bulk_rate_scaling}
\end{figure}

In~\cref{fig:results:bulk_rate_scaling} (left) we demonstrate this behaviour where our method works reliably ($\mathcal{A} \approx 1$) %
for large values of the bulk rates ($\gamma \gg 1$). In fact $\mathcal{A}$ stays close to unity %
until rather low $\gamma \approx 10^{-3}$, and it is only at that point that the performance of the method drops significantly. That this has to be happen is expected since in the limit of {\em slow} bulk reactions ($\gamma \ll 1$) the memory term and the random force both become small and a fixed trajectory length $T$ is not long enough to detect the effect of the bulk. In fact in the limit $\gamma \rightarrow 0$, the subnetwork and bulk become fully decoupled.

Intuitively and from equation~\cref{eq:results:projected_equation_bulk_rate_scaling} we expect that in the small $\gamma$ regime we need long trajectories of length $T\propto 1/\gamma$ to detect the effect of the bulk. For a fixed ``budget'' $N$ of data points for training, such large $T$ then also implies %
large time steps, $\Delta t = \tilde{N} T/N \propto 1/\gamma$, where the subnetwork dynamics can eventually no longer be resolved. In~\cref{fig:results:bulk_rate_scaling} (right) we plot in red the $\mathcal{A}$ %
when we scale both the total trajectory length and the timestep in this way, i.e.\ with $1/\gamma$. %
This does not significantly improve performance over constant $T$ and $\Delta t$: while we have dealt with the issue of $T$ being too small, we now have timesteps $\Delta t$ that are too large when $\gamma$ is small. 
If we instead relax the constraint of fixed $N$ and keep $\Delta t$ small enough to resolve the subnetwork dynamics, then the method works reliably even at small $\gamma$ as can be seen by the orange line in~\cref{fig:results:bulk_rate_scaling} (right). 

In summary, we have shown in this section that, with sufficient temporal resolution of the subnetwork dynamics and an integration time that covers the characteristic relaxation time of the system including the bulk, the boundary detection method also works for dynamics with mixed time scales.

\section{Conservation laws and steady states}
\label{sec:results:conservation_laws}

\begin{figure}[t]
    \centering
    \includegraphics[width=.8\linewidth]{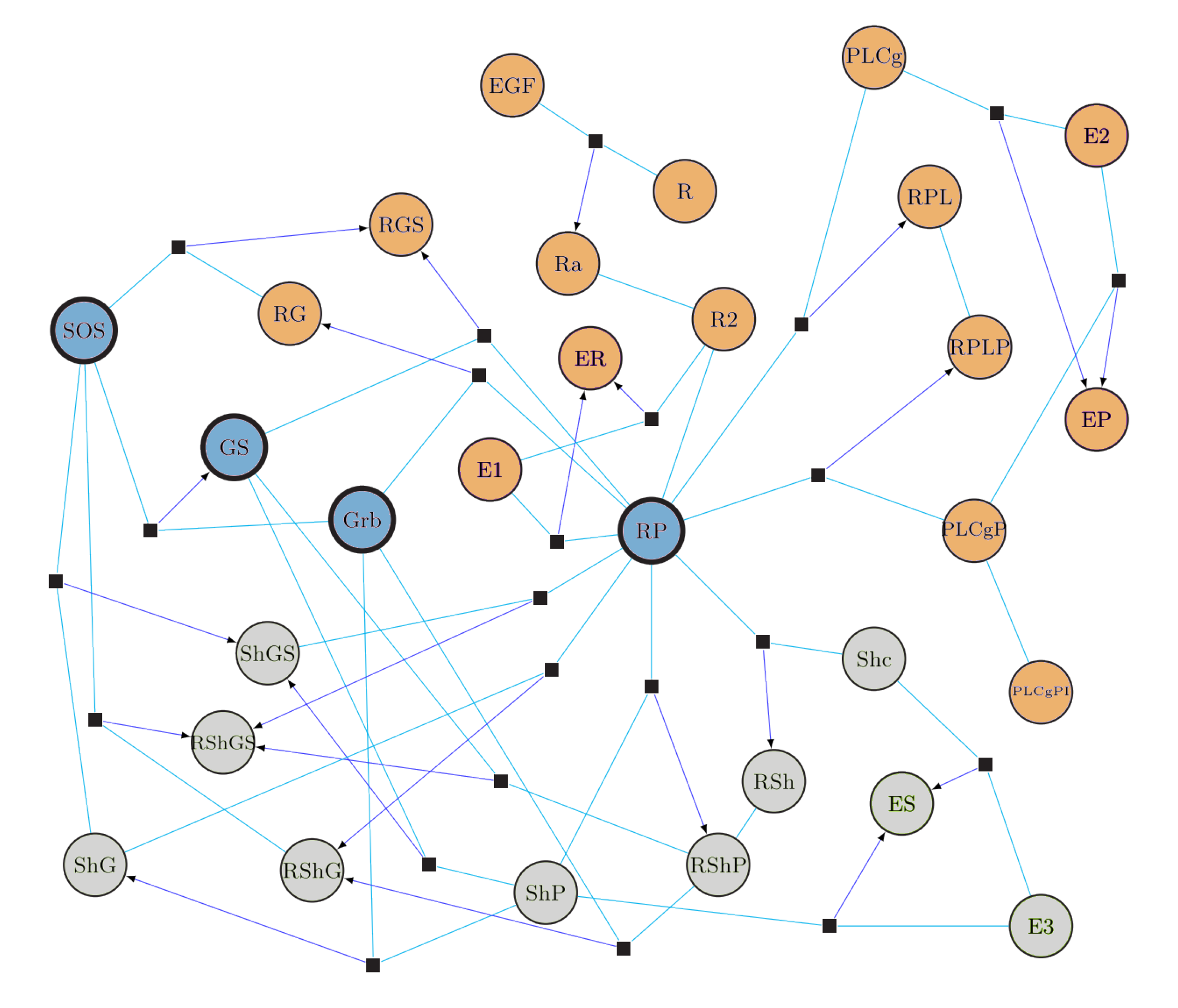}
    \caption{The 29 species and the connectivity of the EGFR network. The thick bordered circles in blue show the boundary species, while the orange ones are the interior species which together form the observed sub-network as opposed to the unobserved bulk species in gray. The enzymes are E1, E2 and E3, with the corresponding substrate-enzyme complexes ER, EP and ES; PLCg, R2 and Shc are the products of the enzymatic reactions from the respective substrates PLCgP, RP and Shp. The enzymes and their complexes are taken as unobserved because we directly use MM dynamics for the conversion of the substrate to the product, as in the original network parameterization~\cite{kholodenko_quantification_1999}.} %
    \label{fig:results:egfr_network_diagram}
\end{figure}

With regard to the number of experiments, $\tilde N$, required for the training of the algorithm, the existence of conservation laws becomes important. Looking back at~\cref{eq:compact_time_evolution}, i.e.\ $ \partial_t \bm x^* = \bm S \bm f$, one sees that any vector $\bm e$ in the left null space of the stoichiometric matrix $\bm S$ corresponds to a conservation law because $\bm e^\text{T} \bm S = 0$ implies $\partial_t (\bm e^\text{T}\bm x^*)=0$.

We are particularly interested in the case when the conservation laws include a combination of boundary and bulk species, such that the bulk species can be expressed in terms of the conserved quantities in the dynamics of the boundary species.

For the simple five species network we can determine the left null space of the stoichiometric matrix explicitly to obtain three conserved quantities:
   \begin{align}
        f_1(\bm x^*) =\ & x_2 + x_3 = c_1 \\
        f_2(\bm x^*) =\ & x_1 - x_2 - x_4 = c_2\\
        f_3(\bm x^*) =\ & x_1 - x_2 + x_5 = c_3
        \label{eq:results:conserved_sums_simple_network}
    \end{align}
We can use these to eliminate $x_2$, $x_4$ and $x_5$ from the dynamics to get the following equation for $x_1$:
    \begin{equation}
        \begin{split}
            \partial_t x_1 = & [-c_1 k_{12,3}   - k_{5,14} + (c_1 + c_2 )k_{14,5}] x_1 + (k_{3,12} - k_{5,14})  x_3 + (k_{12,3} - k_{14,5})  x_1 x_3  \\ 
            &- k_{14,5} x_1^2 + (c_1 + c_3)  k_{5,14} 
            \label{eq:results:dynamical_eq_indep_species}    
        \end{split}
    \end{equation} 
If data from only one experiment -- and thus for a single set of $(c_1,c_2,c_3)$ -- are provided for training, then we see that the dynamics can again be represented in terms of some apparent rate constants. 
If we use our regression algorithm with $\tilde N = 1$ we will thus not see any memory effects, and boundary species will not be detectable.
This analysis shows that larger values of $\tilde N$ are desirable for our detection approach, and more particularly that it is the diversity of steady states reached in the different experiments 
that is helpful for reliable boundary species detection.
We have verified this prediction explicitly by comparing our method on two types of data sets, the first consisting of trajectories with random initial conditions, the second generated specifically to contain only trajectories with identical steady states. As expected, boundary detection performance is higher for the first type (data not shown). %

We have also tested the performance of our method as we change the size of the initial perturbation relative to the steady state, i.e.\ $\delta \bm x_i(0)/\bm y_i$. We find that the performance of the boundary detection method is independent of how small or big the initial perturbation is, as long as the system relaxes to diverse steady states in the $\tilde{N}$ different experiments (data not shown).

\begin{figure}
    \centering
    \includegraphics[width=0.95\textwidth]{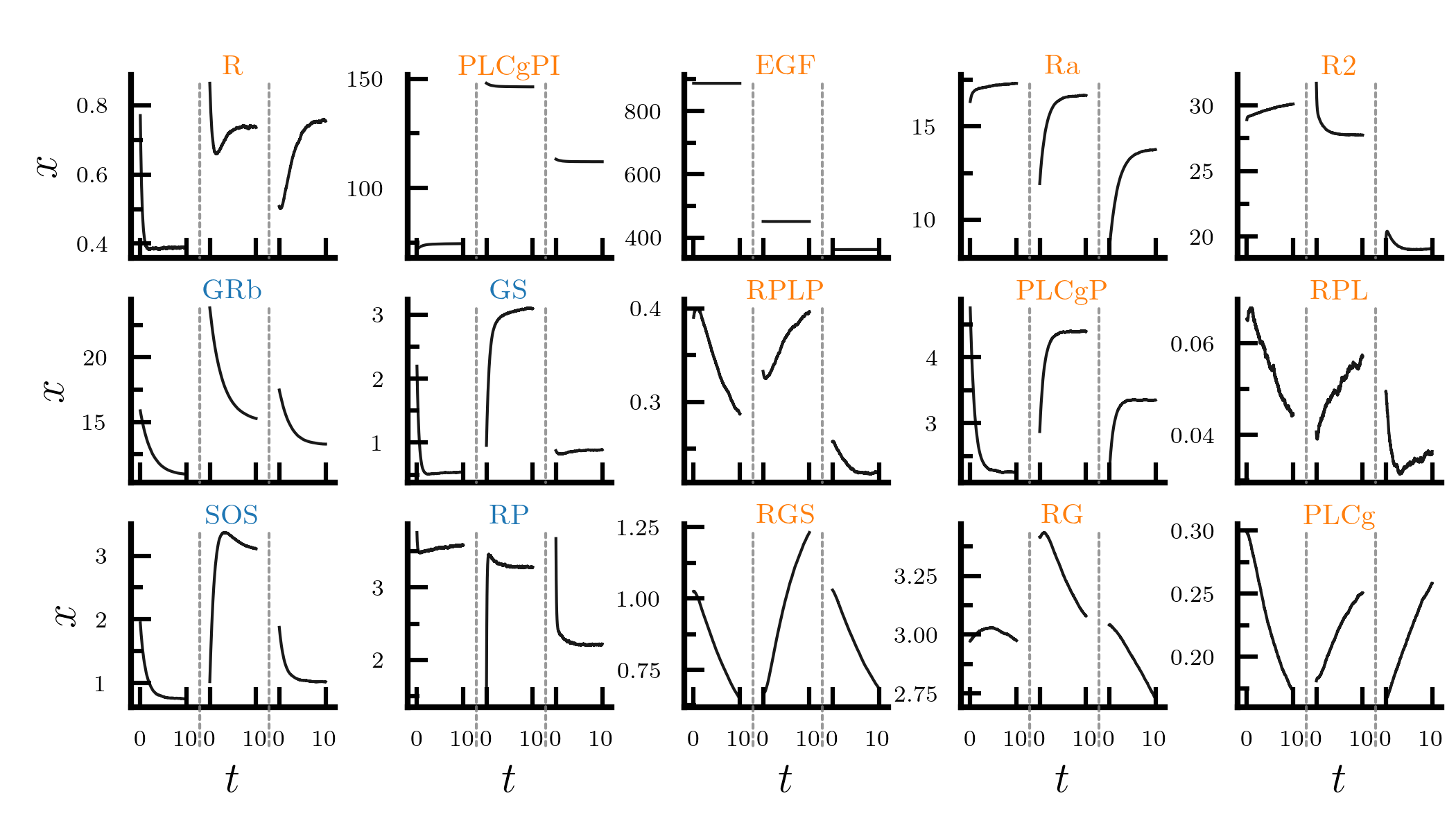}
    \caption{Sample time traces of the EGFR network with concentrations $x$ in nM and $t$ in seconds. The parameters are $\tilde N = 300$, $\epsilon=10^{-5}$ nM,  
    $\Delta t=5\times10^{-3}$ s, $T=10$ s, but we show only three experiments. The four boundary species (SOS, RP, GS, GRb) are shown in blue, while the interior subnetwork species are in orange. %
    }
    \label{fig:egfr_time_traces}
\end{figure}

\begin{figure}
    \centering
    \includegraphics[width=.95\textwidth]{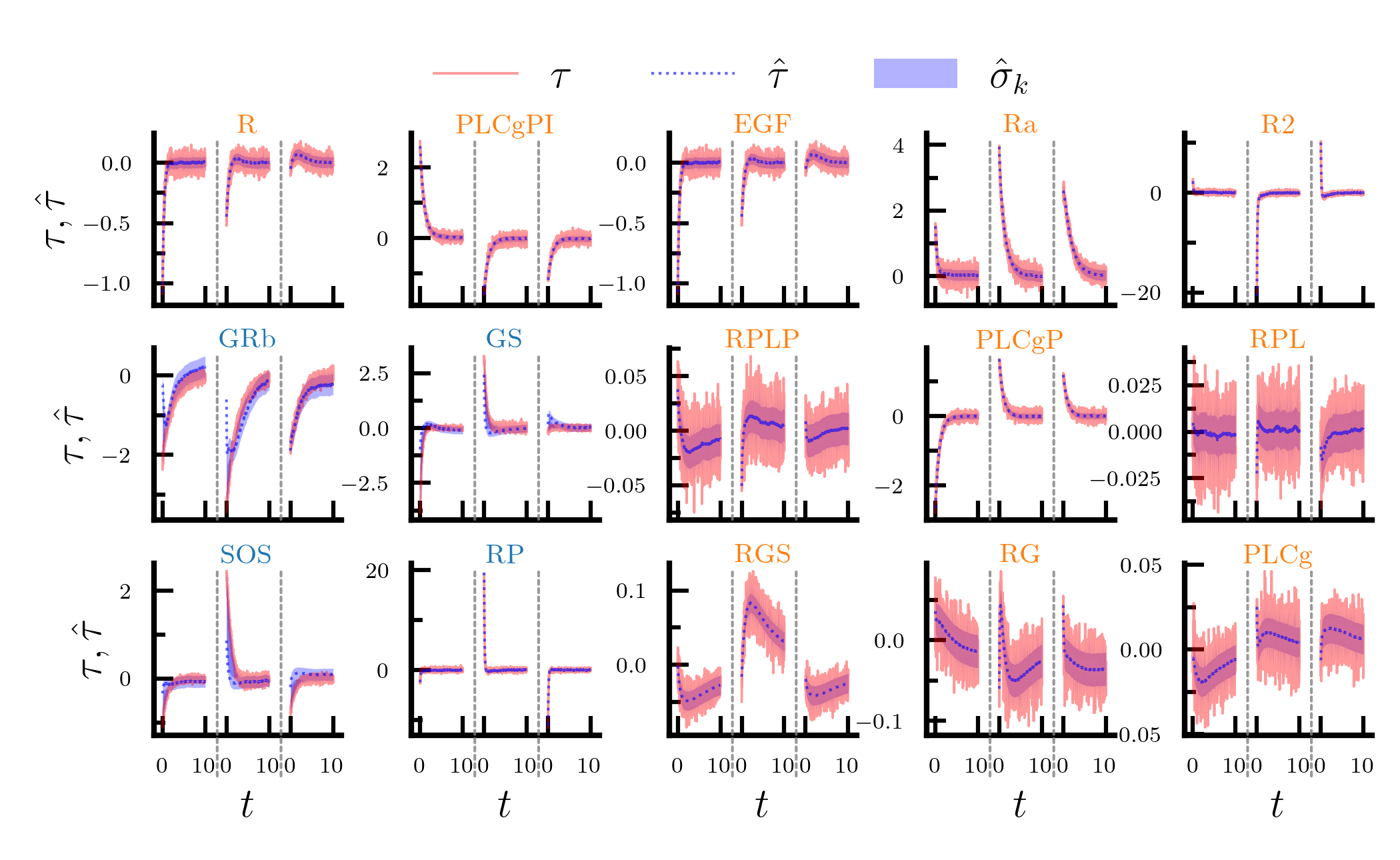}
    \caption{The concentration time derivatives or the regression targets, $\tau$ (pink lines), and their regression estimates, $\hat \tau$ (purple lines), both in units of nM s$^{-1}$, vs $t$ (in seconds). The predicted standard deviation $\hat \sigma$ is shown as the purple shaded area for the boundary (SOS, RP, GS, GRb) and the other interior species. Same parameters as in~\cref{fig:egfr_time_traces}; additionally the posterior (hyper-)parameters $\hat \alpha$, $\hat \beta$ are averaged over three training data sets to generate the posterior distributions. 
    }
    \label{fig:egfr_derivatives_regression}
\end{figure}

\begin{figure}[t]
    \centering
    \includegraphics[width=.95\textwidth]{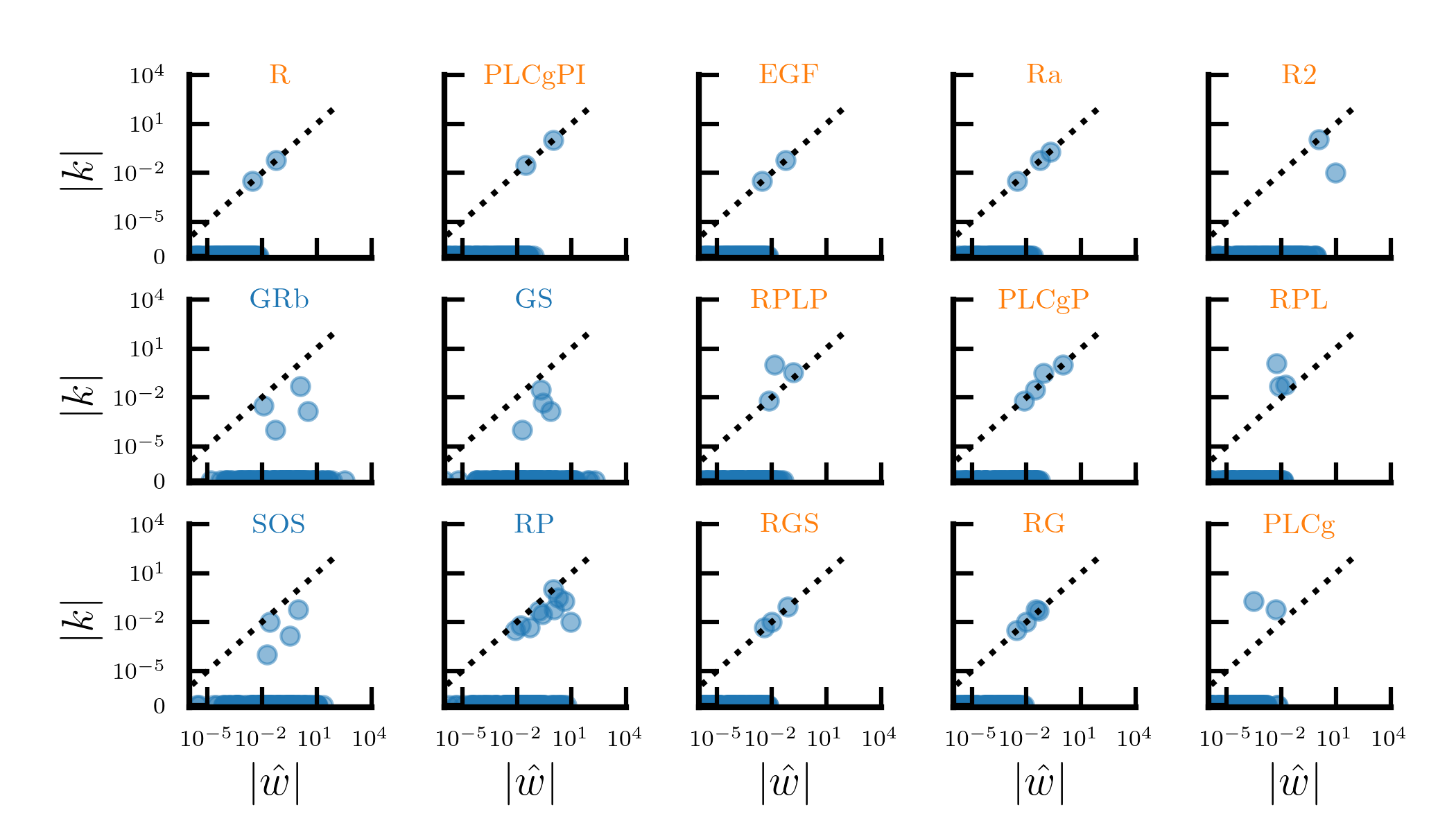}
    \caption{Comparison of the absolute values of the true reaction constants ($|k|$) and their estimates in terms of regression weights ($|\hat w|$) for all the basis functions, for the same parameters as for~\cref{fig:egfr_derivatives_regression} and additionally the regression weights are averaged over three runs. 
    The boundary detection algorithm predicts the weights for the interior species (names in orange) much more reliably (points along the diagonal) than those for the boundary species (names in blue). We include the weights predicted for the basis functions corresponding to the reactions that are absent, i.e.\ have $k=0$, in the original subnetwork. These spurious weights take small values (compared to the actual rate constants) for the interior species but cover a large range of values (comparable to those of actual rate constants) for the boundary species.}

    \label{fig:egfr_weight_estimates}
\end{figure}

\section{EGFR network and effect of noise}
\label{sec:egfr}

We provide here some more detail on the EGFR network to which we apply our boundary detection method in the main text. 
The network topology is shown in~\cref{fig:results:egfr_network_diagram}. Further details can be found in~\cite{kholodenko_quantification_1999}, where the reaction kinetics of the network were quantified including the steady state concentrations (summarized in~\cref{tab:appendix:egfr_steady_state_values_kholodenko}) and the associated rate constants (see~\cref{tab:appendix:egfr_rate_constants_heterodimer,tab:appendix:egfr_rate_constants_unary,tab:appendix:egfr_michelis_menten_konstants}). What is notable is that in both cases there is a wide variation, across some four orders of magnitude. 

Out of the 29 species we take 19 species as our observed subnetwork, with SOS, GS, Grb and RP constituting the boundary species. We simulate the system using Michaelis-Menten (MM) kinetics for the enzyme reactions, and thus do not include explicit observations of enzymes or enzyme complexes. Leaving out these three enzymes and three enzyme complexes, we have 23 species in the EGFR network. Two enzyme and two enzymatic complexes were in the observed subnetwork and excluding them leaves 15 species in the observed subnetwork.

We show illustrative noisy time traces in~\cref{fig:egfr_time_traces} for the concentrations of the different species. \Cref{fig:egfr_derivatives_regression} has the concentration time derivatives, i.e.\ the regression targets, the regression fits and the predictive standard deviations obtained from the regression for each molecular species. The parameters for the inference from the noisy traces for these plots and those in the main paper are $\tilde N = 300$, $\epsilon=10^{-5}$\,nM,  
$\Delta t=5\times10^{-3}$\,s, $T=10$\,s.

\Cref{fig:egfr_weight_estimates} displays an example of regression estimates of the weights of the subnetwork species, plotted against the true reaction constants. We see that, for the interior species, the algorithm predicts the true reaction constants (points along the diagonal), along with some small spurious weights. For the boundary species, these spurious weights are larger and the true rate constants are not recovered, both as expected.

\begin{figure} \centering
    \includegraphics[width=.8\linewidth]{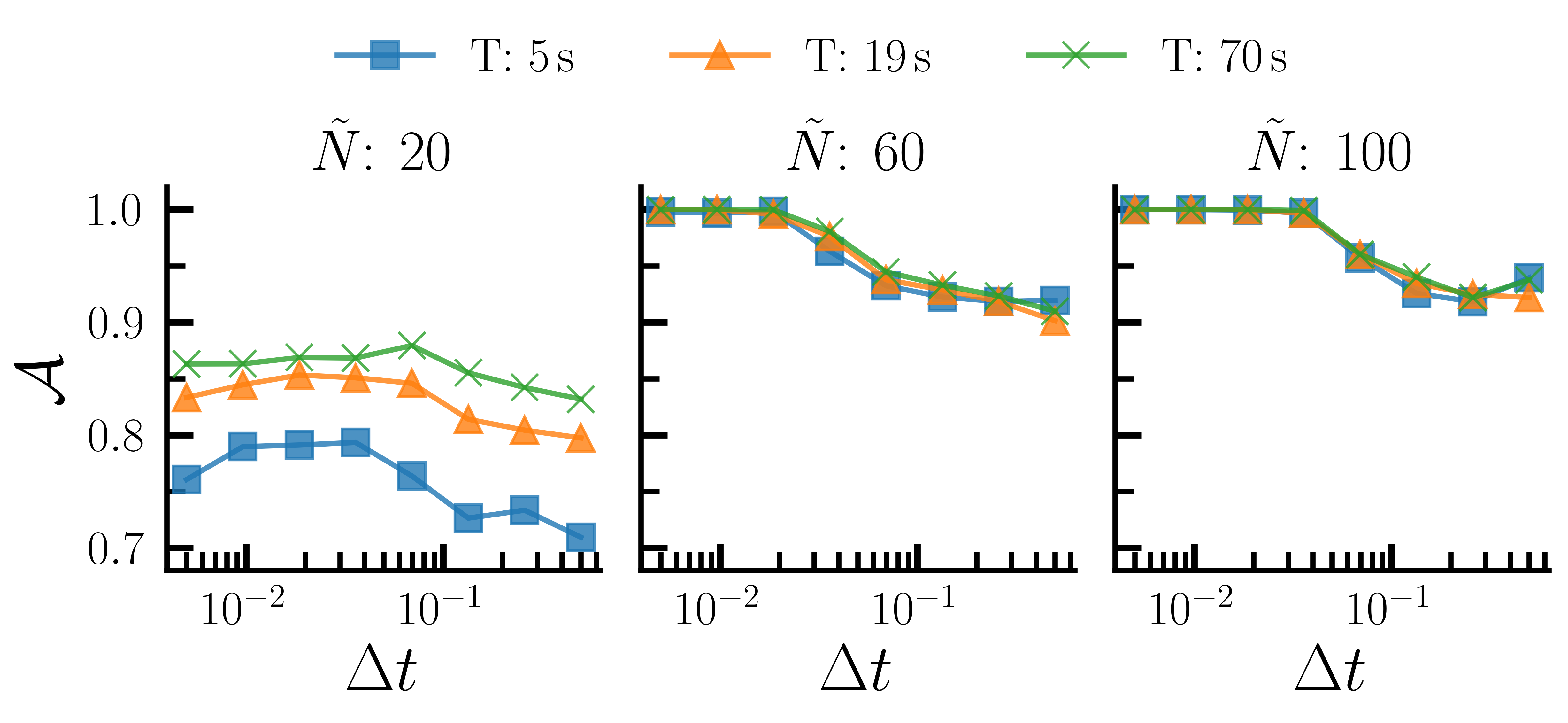}
    \caption{ The AUC $\mathcal{A}$, for the noise free EGFR network for different $\tilde N$ and $T$ values as we change $\Delta t$ (in seconds). 
    The optimal parameter regime is $\tilde N \geq 60$, $T \geq 5$\,s, $\Delta t \leq 0.05$\,s.} %
    \label{fig:parameter_egfr_noise_free_nexp}
\end{figure}

In the noise free case, we again evaluate the performance of our method for the EGFR network as we change $\tilde N,\ T \text{ and } \Delta t$, see~\cref{fig:parameter_egfr_noise_free_nexp}. We find that we need a larger number of independent experiments, $\tilde N \geq 60$, with small $\Delta t \leq 0.05$\,s and $T \geq 5$\,s to resolve the typical relaxation dynamics. Note that some relaxations in the EGFR network happen on much longer time scales of $100$\,s, but it turns out that our approach remains viable nonetheless with significantly smaller $T$.

For noisy data it is useful to scale the inferred effective noise variances by an estimate of the stochastic variance $\sigma^2_{\text{emp}}$, as explained in the main text. For EGFR we measured $\sigma^2_{\text{emp}}$ from the initial non-stationary part of the time courses instead of the steady state, by calculating the variance of a noisy trajectory around the noiseless trajectory with the same parameters. The reason for this was that many species of the EGFR network have very low steady state concentrations that are difficult to simulate using the boundary reflection algorithm for the Langevin equation. This issue would obviously not arise for experimental biochemical measurements.

\begin{table}[]
    \centering
    \begin{minipage}[t!]{0.3\textwidth}
    \begin{tabular}{rr}
    \multicolumn{1}{l}{\textbf{species}} & \multicolumn{1}{l}{\textbf{s.s.}} (nM) \\ \hline
    \rowcolor[HTML]{EFEFEF} 
    EGF                                  & 580.60                                    \\
    GRb                                  & 14.17                                     \\
    \rowcolor[HTML]{EFEFEF} 
    GS                                   & 1.70                                      \\
    PLCg                                 & 0.18                                      \\
    \rowcolor[HTML]{EFEFEF} 
    PLCgP                                & 3.04                                      \\
    PLCgPI                               & 101.42                                    \\
    \rowcolor[HTML]{EFEFEF} 
    R                                    & 0.60                                      \\
    R2                                   & 30.30                                     \\
    \end{tabular}
    \end{minipage}
    \begin{minipage}[t!]{0.3\textwidth}
    \begin{tabular}{rr}
    \multicolumn{1}{l}{\textbf{species}} & \multicolumn{1}{l}{\textbf{s.s.}} (nM) \\ \hline
    \rowcolor[HTML]{EFEFEF} 
    RG                                   & 3.07                                      \\
    RGS                                  & 0.91                                      \\
    \rowcolor[HTML]{EFEFEF} 
    RP                                   & 3.60                                      \\
    RPL                                  & 0.05                                      \\
    \rowcolor[HTML]{EFEFEF} 
    RPLP                                 & 0.32                                      \\
    RSh                                  & 0.07                                      \\
    \rowcolor[HTML]{EFEFEF} 
    RShG                                 & 0.48                                      \\
    RShGS                                & 0.34                                      \\
    \end{tabular}
    \end{minipage}
    \begin{minipage}[t!]{0.3\textwidth}
    \begin{tabular}{rr}
    \multicolumn{1}{l}{\textbf{species}} & \multicolumn{1}{l}{\textbf{s.s.}} (nM) \\ \hline
    \rowcolor[HTML]{EFEFEF} 
    RShP                                 & 1.84                                      \\
    Ra                                   & 17.41                                     \\
    \rowcolor[HTML]{EFEFEF} 
    SOS                                  & 1.78                                      \\
    ShG                                  & 35.07                                     \\
    \rowcolor[HTML]{EFEFEF} 
    ShGS                                 & 29.28                                     \\
    ShP                                  & 81.78                                     \\
    \rowcolor[HTML]{EFEFEF} 
    Shc                                  & 1.15                                      \\
                                         & 
    \end{tabular}
    \end{minipage}
    \caption{Steady state (s.s.) values for the EGFR network.}
    \label{tab:appendix:egfr_steady_state_values_kholodenko}
    \end{table}

    \begin{table}
    \centering
    \begin{tabular}{rrrrr}
    \multicolumn{1}{l}{\textbf{educt 1}} & \multicolumn{1}{l}{\textbf{educt 2}} & \multicolumn{1}{l}{\textbf{product}} & \multicolumn{1}{l}{\textbf{rate forward}} $(\text{nM}^{-1}\ \text{s}^{-1})$ & \multicolumn{1}{l}{\textbf{rate backward}}$(\text{s}^{-1})$ \\ \hline
    \rowcolor[HTML]{EFEFEF} 
    GRb                                  & SOS                                  & GS                                   & 0.00010                                   & 0.0015                                     \\
    RP                                   & GRb                                  & RG                                   & 0.00300                                   & 0.0500                                     \\
    \rowcolor[HTML]{EFEFEF} 
    RG                                   & SOS                                  & RGS                                  & 0.01000                                   & 0.0600                                     \\
    RP                                   & GS                                   & RGS                                  & 0.00450                                   & 0.0300                                     \\
    \rowcolor[HTML]{EFEFEF} 
    RP                                   & PLCg                                 & RPL                                  & 0.06000                                   & 0.2000                                     \\
    RP                                   & PLCgP                                & RPLP                                 & 0.00600                                   & 0.3000                                     \\
    \rowcolor[HTML]{EFEFEF} 
    RP                                   & Shc                                  & RSh                                  & 0.09000                                   & 0.6000                                     \\
    RShP                                 & GRb                                  & RShG                                 & 0.00300                                   & 0.1000                                     \\
    \rowcolor[HTML]{EFEFEF} 
    RP                                   & ShG                                  & RShG                                 & 0.00090                                   & 0.3000                                     \\
    RShG                                 & SOS                                  & RShGS                                & 0.01000                                   & 0.0214                                     \\
    \rowcolor[HTML]{EFEFEF} 
    RShP                                 & GS                                   & RShGS                                & 0.00900                                   & 0.0429                                     \\
    ShGS                                 & RP                                   & RShGS                                & 0.00024                                   & 0.1200                                     \\
    \rowcolor[HTML]{EFEFEF} 
    ShP                                  & RP                                   & RShP                                 & 0.00090                                   & 0.3000                                     \\
    R                                    & EGF                                  & Ra                                   & 0.00300                                   & 0.0600                                     \\
    \rowcolor[HTML]{EFEFEF} 
    ShP                                  & GRb                                  & ShG                                  & 0.00300                                   & 0.1000                                     \\
    ShG                                  & SOS                                  & ShGS                                 & 0.03000                                   & 0.0640                                     \\
    \rowcolor[HTML]{EFEFEF} 
    ShP                                  & GS                                   & ShGS                                 & 0.02100                                   & 0.1000                                     \\
    Ra             & Ra                   & R2               & 0.01                  & 0.1 \\
    \end{tabular}
    \caption{Rate constants for the heterodimer reactions and the homodimer reaction (the last row) in the EGFR network.} %
    \label{tab:appendix:egfr_rate_constants_heterodimer}
\end{table}

\begin{table}
    \centering
    \begin{tabular}{rrrr}
   \textbf{educt} & \textbf{product} & \textbf{rate forward} $(\text{s}^{-1})$ & \textbf{rate backward} $(\text{s}^{-1})$ \\ \hline
    \rowcolor[HTML]{EFEFEF} 
    PLCgP          & PLCgPI           & 1                     & 0.03                   \\
    R2             & RP               & 1                     & 0.01                   \\
    \rowcolor[HTML]{EFEFEF} 
    RPL            & RPLP             & 1                     & 0.05                   \\
    RSh            & RShP             & 6                     & 0.06                  
    \end{tabular}
    \caption{Rate constants for (non-enzymatic) unary reactions of the EGFR network.} 
    \label{tab:appendix:egfr_rate_constants_unary}
    \end{table}
\begin{table}
    \centering
    \begin{tabular}{rrrr}
    \textbf{substrate} & \textbf{product} & $\bm {V}_\text{m}$ (nM\ s$^{-1}$) & {$\bm {K}_\text{m}$ (nM)} \\ \hline
    \rowcolor[HTML]{EFEFEF} 
    PLCgP              & PLCg             & 1.0                           & 100                  \\
    RP                 & R2               & 450.0                         & 50                   \\
    \rowcolor[HTML]{EFEFEF} 
    ShP                & Shc              & 1.7                           & 340                 
    \end{tabular}
    \caption{Michaelis-Menten constants $V_m, K_m$ for the enzymatic reactions in the EGFR network. %
    }
    \label{tab:appendix:egfr_michelis_menten_konstants}
    \end{table}
    
    \clearpage
    \printbibliography